\documentclass[%
 aip,
 amsmath,amssymb,
 reprint,%
]{revtex4-1}

\usepackage{graphicx}
\usepackage{dcolumn}
\usepackage{bm}

\usepackage[utf8]{inputenc}
\usepackage[T1]{fontenc}
\usepackage{mathptmx}
\usepackage{xspace}

\begin{document}

\preprint{AIP/123-QED}

\title[]{Supercell-core software: a useful tool to generate an optimal supercell for vertically stacked nanomaterials\\}

\author{Tomasz Necio}
\author{Magdalena Birowska}
\email{Magdalena.Birowska@fuw.edu.pl}
 \affiliation{University of Warsaw, Faculty of Physics, 00-092 Warsaw, Pasteura 5, Poland}


\begin{abstract}
Vertically oriented materials, such as van der Waals heterostructures, that have novel hybrid properties are crucial for fundamental scientific research and the design of new nano-devices. Currently, most available theoretical methods require applying a supercell approach with periodic boundary conditions to explore the electronic properties of such nanomaterials. Herein, we present \texttt{supercell-core} software, which provides a way to determine the supercell of non-commensurate lattices, in particular, van der Waals heterostructures. Although this approach is very common, most of the reported work still uses supercells that are constructed “by hand” and on a temporary basis. The developed software is designed to facilitate finding and constructing optimised supercells (i.e., with small size and minimal strain accumulation in adjacent layers) of vertically stacked lattices. 
\end{abstract}

\maketitle
\section{\label{sec:Intr}Introduction
}

When a 2D crystal is placed on top of another layer (substrate) it can either adjust its position (e.g., by rotating to follow the periodic potential of the substrate), resulting in a commensurate state, or the layers can exhibit a small lattice mismatch, in which case, the interlayer binding energy (via weak van der Waals forces) can compensate for the increase in the elastic energy. Both of these occurrences may lead to experimentally observed superperiodic structures, which are commonly known as moir\'e patterns. Such moir\'e superlattices have been widely studied in the context of many different systems, including bilayer graphene \cite{PhysRevB.78.125406,PhysRevLett.109.186807,Hunt1427,Cao2018ins}, trilayer graphene \cite{PhysRevB.101.224107}, and bilayer black phosphorus \cite{PhysRevB.96.195406}, as well as for many heterostructures, such as graphene on hexagonal boron nitride (hBN) \cite{Decker2011,Woods2014}, or  bilayer transition metal dichalcogenides (TMDC) \cite{Kang2013,Alexeev2019}.

 Hybrid structures consisting of different types of layers connected via weak van der Waals forces represent a new class of hybrid crystals known as van der Waals heterostructures \cite{Geim2013}. These structures are of broad interest to many researchers throughout the world, due to the novel hybrid properties arising in these materials, which are distinct from their individual layer components \cite{NatureJENS,Cao2018,Birowska_2019}. Many of these properties can be precisely controlled by rotating the two stacked atomic layers with respect to one another \cite{Cao2018, Cao2018ins,PhysRevB.96.195406}, highlighting the fact that manipulating this unique "twist angle" degree of freedom can allow for control of the nanoscale properties of such nanomaterials \cite{PhysRevB.95.075420,Cao2018}.

The strain itself is the subject of a new research field in solid state physics, called straintronics \cite{Bukharaev2008}, in which strain engineering methods are used to develop next-generation devices for information, sensor, and energy-saving technologies. The strain-induced physical effects in nanolayers or heterostructures, such as changes in the band structure, or electronic, optical, or magnetic properties \cite{C7CP03558F,TAMLEH2018339,AIPMilowska}, are fundamentally important. Understanding these relationships is a prerequisite for developing new technologies using such nanomaterials. The strain distribution can be investigated experimentally using a variety of experimental methods \cite{Woods2014,Xue2011}, such as atomic force microscopy (AFM), scanning tunnelling microscopy (STM), and/or Raman spectroscopy.

Proper modelling is required in order to better understand the phenomena and underlying physics behind the structure-properties relationships. The modelling of this type of nanomaterials mostly focuses on electronic band structure calculations, and one of the most accurate \textit{ab initio} methods is based on density functional theory (DFT). Widely-used software packages, such as VASP \cite{KRESSE199615,PhysRevB.47.558}, Quantum Espresso \cite{QE}, and SIESTA \cite{Soler_2002}, are limited in terms of their supercell approaches, because only a few hundred atoms can be considered. However, stacking adjacent layers with different lattice parameters or different lateral crystal symmetries may allow appropriate modelling of superperiodicty in very large lateral cells, or detection of aperiodic structures. Incommensurate lattices are outside the scope of the present paper. One approach directed towards describing aperiodic layered structures can be found in reference \cite{PhysRevB.101.224107}. In the current work, we discuss a developed approach based on commensurate lattices. We propose a supercell software capable of finding the optimal supercell, i.e., those with small size and low strain distribution. Specifically, we have developed a software package that searches through all possible superperiodicities arising from multiples of primary cells for a given rotation angle between the top and bottom 2D lattices. The software allows determination of the optimal "magic angles"  between the adjacent vertically-stacked layers and the resulting moir\'e patterns.

There are few available builders that can handle the construction of supercells. To the best of our knowledge, those which are freely available, such as VESTA \cite{VESTA}, only enable \textit{ad hoc} construction by hand, without finding the optimal supercell. The other capable programs, such as Quantum Wise \cite{smidstrup2020quantumatk}, which has an appropriately implemented builder, must be purchased. Herein, we  present the package, \texttt{supercell-core}, which is free of charge and allows the user to find the optimal supercell for a given twist angle and "n" number of layers constituting the van der Waals structures. To the best of our knowledge, none of the above mentioned codes, calculate the strain distribution of adjacent layers for a particular angle between the layers.

In this article, we first present the mathematical background for the software, along with explanations of the applied algorithms and discussion of relevant technical details. Then, the code is briefly described, focusing on its functionalities. Finally, we present the practical applications of the software and provide several examples.

\section{\label{sec:details}{Computational methodology}}

\subsection{Mathematical description}

Our methodology is based on a commensurability condition, which requires long-range order in the sets of lattice planes at the interfaces of $n$ vertically stacked layers. 

\begin{figure}
    \centering
    \includegraphics[width=\linewidth]{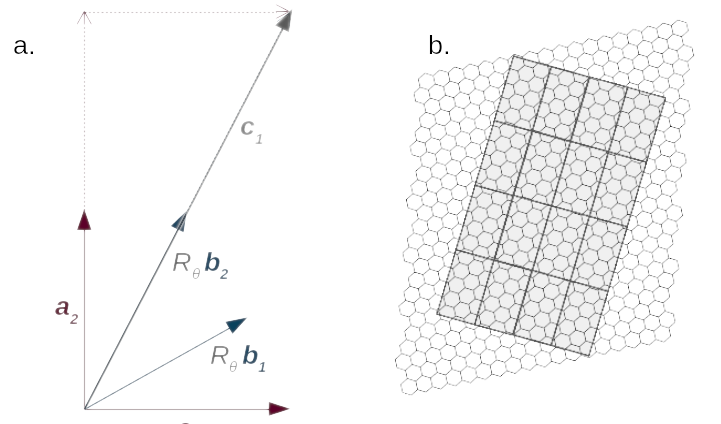}
    \caption{a. Schematic construction of a superlattice vector, $\mathbf{c}_1 = 2 \mathbf{a}_2 + \mathbf{a}_1 = 2 R_\theta \mathbf{b}_2$. b. Example of two incommensurate lattices (hexagonal and rectangular).}
    \label{fig:c_vector}
\end{figure}

In order to formulate this condition mathematically, we can consider two planar lattices: A (bottom layer; substrate) and B (top layer), where lattice B is placed on top of lattice A. 
We denote the unit vectors of both lattices as $\mathbf{a}_i$ and $\mathbf{b}_i$, respectively, where $i \in \{1, 2\}$. The task is to find two new vectors, $\mathbf{c}_i$, that are commensurate to both $\mathbf{a}_i$ and $\mathbf{b}_i$ \cite{taksiazka}, such that,
\begin{equation}
    \exists \, m_{ij} \in \mathbb{Z} \, \land \, \exists \, n_{ij} \in \mathbb{Z}: \; \mathbf{c}_i = \sum_{j=1}^2 (m_{ij})^T \mathbf{a}_j = \sum_{j=1}^2 (n_{ij})^T R_\theta \mathbf{b}_j,
\end{equation}
where $R_\theta$ is the 2D rotation matrix based on the rotation angle between layers, $\theta$ (Fig. \ref{fig:c_vector}a). Depending on the specific case, $\theta$ might be fixed, or it might be a free parameter. The angle, $\theta$, is hereafter referred to as the twist angle.

{}

In principle, there may be no set of integers, $m_{ij}$, $n_{ij}$, which satisfies both of the above equations, even if $\theta$ is treated as a free parameter (Fig. \ref{fig:c_vector}b). Thus, the commensurability condition is enforced by applying strain to the top layer. This corresponds to a linear transformation of the B lattice's elementary cell. Note that layer A (the substrate) is always unstrained. The new task is to find the linear transformation that introduces minimal strain into the system.

For now, let us consider a fixed value of $\theta$. We denote the modified vectors of the rotated lattice B as $\mathbf{\tilde{b}}_i$. Then, we can define its strain tensor, $\varepsilon$, as a 2D matrix:
\begin{equation}
 \sum_{j=1}^2 (\varepsilon + I)^{ij} R_\theta \mathbf{b}_j = \mathbf{\tilde{b}}_i. 
 \end{equation}
The $m_{ij}$ coefficients can easily be calculated given the $n_{ij}$ and $\mathbf{\tilde{b}}_j$ values. We must find  a matrix, $\tilde{B}$, for every set of four numbers, $n_{ij}$, such that $\tilde{B}(\mathbf{b_i}) = \mathbf{\tilde{b}}_i$. Then, from all sets of $n_{ij}$, we choose the one with minimal strain. To quantify the strain, we define the norm $L_{(1, 1)}(\varepsilon)$, which is equal to $\sum_{ij} |\varepsilon_{ij}|$.

Note that, to find a nearly unstrained supercell for a given value of $\theta$, the software searches all possible values of $n_{ij}$ (up to some maximum value, defined by the \texttt{max\_el} parameter in the code) in order to be certain that the optimal supercell is determined. However, this approach (hereafter referred to as the \texttt{Direct} algorithm), has a time complexity of $O(N^4)$, where $N$ is the upper limit of $n_{ij}$ that is considered. However, the problem of finding a nearly unstrained supercell can be greatly simplified. It is important to note that, when the strain is zero, then $\mathbf{b}_i = \mathbf{\tilde{b}}_i$ (see Eq. 2). For every pair of $n_{i1}, n_{i2}$, it is possible to write independent linear equations for $m_{i1}, m_{i2}$:

\begin{equation} \label{eq1}
    \begin{pmatrix}
    \mathbf{a}_1 & \mathbf{a}_2
    \end{pmatrix}^{-1}
    \begin{pmatrix}
        \cos\theta & -\sin\theta \\
    \sin\theta & \cos\theta
    \end{pmatrix}
    \begin{pmatrix}
    \mathbf{b}_1 & \mathbf{b}_2 
    \end{pmatrix}
    \begin{pmatrix}
    m_1 \\ m_2
    \end{pmatrix}
    =
    \begin{pmatrix}
    n_1 \\ n_2
    \end{pmatrix}
\end{equation}

\begin{figure}
\centering
\includegraphics[width=0.5\textwidth]{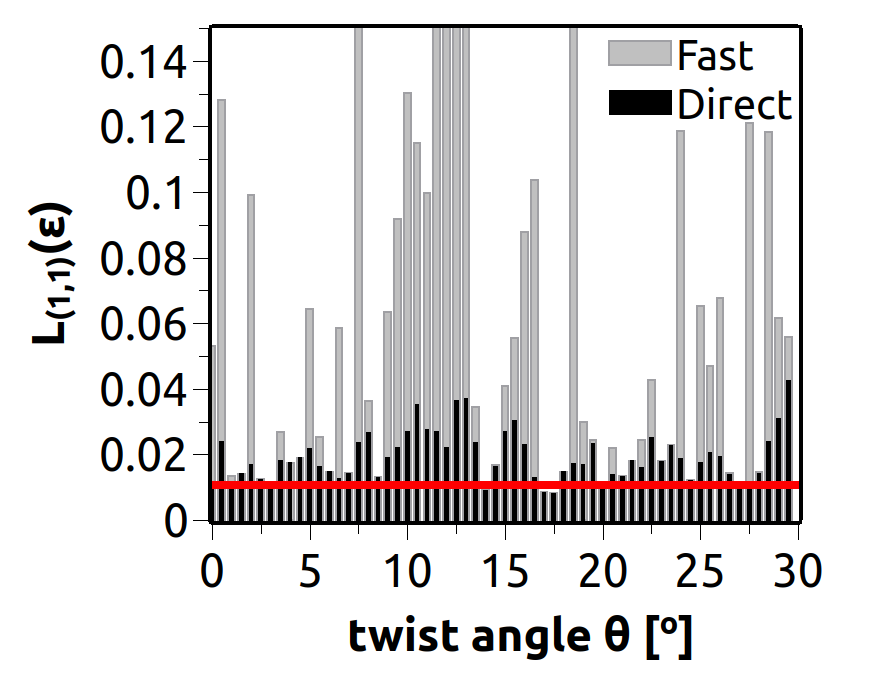}
\caption{Comparison of the Fast and Direct algorithms implemented in the \texttt{supercell-core} software. The norm of the  strain tensor is calculated for an optimal supercell for a phosphorene$/$graphene heterostructure as a function of the twist angle, $\theta$. The red line indicates the value of $L_{(1, 1)}(\varepsilon)$=0.015, for which both algorithms give the same results. Note that, for the highly-strained lattices, the discrepancy between the algorithms can be substantial, and in those cases, use of the \texttt{Direct} algorithm is recommended.}
    \label{fig:por}
\end{figure}

The "quality" of each $n_{ij}$ pair can be assessed by taking a measure of how far the corresponding $m_{ij}$ value is from an integer solution: $\frac{\mathbf{v} - \mathrm{round}(\mathbf{v})}{||\mathbf{v}||}$, where $\mathbf{v} =     \begin{pmatrix}
    m_1\\
    m_2
    \end{pmatrix}$. 
This algorithm, known as the \texttt{Fast} algorithm, effectively describes $O(N^2)$ in time complexity, and gives the same results as a \texttt{direct} search for cases where an optimal supercell's strain is zero or near-zero. It can be used to find the moir\'e patterns or nearly unstrained unit cells. However, the \texttt{Fast} algorithm does not guarantee an optimal solution for cases where an optimised supercell is significantly strained. Our analysis indicates that both algorithms give the same results when the norm $L_{(1, 1)}(\varepsilon)$ does not exceed a value of 0.015 (see Fig. \ref{fig:por}). Thus, it guarantees that the strain calculated with the
\texttt{Fast} algorithm is not overestimated.

These problems trivially extend to cases where $\theta$ is a free parameter, by treating it as an additional degree of freedom when searching the parameter space. When considering multiple layers, B$_1$, B$_2$, etc., we can exploit the fact that all lattices must ultimately be commensurate, and we do not modify lattice A at all. Effectively, each layer can be fit to the same lattice A supercell. Therefore, we can repeat most of the above steps for each layer independently, and record the calculated "qualities" of every set of $m_{ij}$ values for each layer. To select the final supercell (which is determined by its $m_{ij}$ coefficients), we then use the criterion of the lowest sum of the norm $L_{(1, 1)}(\varepsilon)$ for all B$_i$ layers.

\subsection{Implementation}

The code is implemented as a Python package using the Python numerical library, NumPy \cite{SciPy}. Every 2D crystal and every heterostructure is represented by a Python object. Optimisation of the strain for the given values of the twist angle, $\theta$, for each layer is carried out by first preparing a list of all possible combinations of those values. Then, for each combination, an array of all possible vectors of the supercell in the substrate lattice vectors basis is prepared, which generates an $N\times N\times 2$ array, where each two-dimensional vector stored in that array is a grid point in the substrate lattice vector basis. The value of $N$ describes the extent to which we check $N = 2n + 1$ (where $n$ is the highest integer that can be accepted as the $n_{ij}$ coefficient). Certain basic symmetries are utilised to decrease the number of required calculations (decreasing the number of $n_{ij}$ values, e.g., the symmetry with respect to interchanges $\overrightarrow{c_1}$ and $\overrightarrow{c_1}$, rotation about 180$^{\circ}$). 

In the \texttt{Fast} algorithm ("fast" option), the strain tensor is calculated for all possible $\mathbf{c}_i$ vectors, and the $L_{(1, 1)}(\varepsilon)$ is analysed to assess the quality of the specific configuration in a vectorised operation. Ultimately, for a given combination of interlayer angles, the configuration that gives the lowest strain is chosen. In the minimisation process, smaller, less narrow cells are treated preferentially. If the \texttt{Direct} algorithm is applied by the user, the procedure follows the mathematical description of our solution for cases in which the desired strain is near-zero (see Eq. (3)).

For drawings, the code uses the MatPlotLib Python library \cite{SciPy}. An optional dependency of the library is pandas \cite{pandas}, which is applied if the user wants to \texttt{log} intermediate steps of the strain minimisation procedure.

\section{Description of the code}\label{description}

The software is available in the official Python package index as \texttt{supercell\_core}. Assuming that a distribution of Python software is installed, a user can download the package with the command: \texttt{pip install supercell\_core}. The recommended way to use the package in a Python script, Python console, or Jupyter notebook is via an \texttt{import supercell\_core as sc} statement. The source code is also available from \cite{github} in the form of a Git repository. The repository contains examples files, described in the next section in the \texttt{supercell\_core/examples} subdirectory, and a README file with usage examples and a link to online documentation. The package documentation is also provided interactively in the Python console via Python's built-in \texttt{help} function.

The software enables the user to find the optimal configuration for an arbitrary number of vertically stacked layers, and for any set of acceptable interlayer angles between them. The parameter, $L_{\max}(A) = \max |m_{ij}|$, can be used to control the trade-off between computational cost, the quality of the resulting supercell, and its size (\texttt{max$\_$el} in the code, referrers to the substrate layer denoted as  A). \texttt{Supercell-core} accepts definitions of the lattices either performed manually using the \texttt{lattice} function, or read from a VASP POSCAR file with the \texttt{read\_POSCAR} command. For convenience, users are encouraged to use \texttt{supercell-core} with a Jupyter notebook or an interactive Python console.

The user can control the positions of the layers in the $z$-direction by changing the $z$-components of the unit cell vectors of the layers. Note, that the strain distribution is only calculated in lateral coordinates. This software stacks unit cells of the layers directly on top of each other. For example, if one layer with unit cell vectors $(a, b, 0), (c, d, 0), (0, 0, h)$ contains an atom at position $(x, y, z_1)$, and the layer directly above it contains an atom at $(x, y, z_2)$, then the distance between these atoms in the calculated heterostructure will be $h - z_1 + z_2$.

It is important to note that the inclusion of different stacking configurations in primary cells of the layers can be achieved by translating the atoms via a corresponding vector in one of the cells. The described approach constructs the supercell independent of the stacking configurations of the layers, since it is only based on the lattice vectors. Thus, the atomic positions are correspondingly scaled and translated in a new supercell. 

Supercell optimisation and strain calculation methods can be performed on a \texttt{Heterostructure} object. Choices related to the optimisation algorithm, the maximum $n_{ij}$ value, and whether to store intermediate results for each combination of interlayer angles in a \texttt{log}, are conducted by specific arguments in the \texttt{opt} method. All results of these calculations are stored in \texttt{Result} objects. Saving results to XCrysDen XSF and/or VASP POSCAR files is possible using methods for the \texttt{Lattice} object. Every \texttt{Result} object contains a \texttt{superlattice} method, which returns a \texttt{Lattice} corresponding to the heterostructure (containing the correct positions of atoms in all layers of the new supercell). If \texttt{logging} is enabled, the \texttt{log} can be saved to a CSV file.  When this option is used, for each twist angle the code lists in the order the following details: twist angles "\texttt{theta\_0}" etc. in radian unit (at the end in $^{\circ}$); the sum of norm $L_{(1, 1)}(\varepsilon)$ "\texttt{max$\_$strain}"; lateral size in \AA$^2$ "\texttt{supercell\_size}"; matrices describing the vectors of the supercell on the basis of substrate and Cartesian coordinates denoted as "M", "N" etc.; "\texttt{supercell$\_$vectors}" ( $\vec{c_1}$=(supercell$\_$vectors$\_$11, supercel$\_$vectors$\_$21), $\vec{c_2}$=(supercell$\_$vectors$\_$12, supercel$\_$vectors$\_$22) given in \AA); strain tensor $\varepsilon_{ij}$ denoted as "\texttt{strain$\_$tensor$\_$layer}" for each layer B$_i$ and number of atoms "\texttt{atom$\_$count}". If there are multiple layers, the user obtains a table with a row for every combination of twist angles. The list takes the form of a \texttt{pandas.DataFrame} file, which makes it easy to view, transform, and save the data (e.g., to CSV format).

The package is also able to draw the positions of atoms in the supercell (projected onto an xy plane) using the Matplotlib plotting library.

\section{Examples}

Here, we present three examples that demonstrate the capabilities of the developed software and verify the correctness of the obtained results. Specifically, A. bilayer graphene (BG), B. trilayer graphene (TG) and  hBN/graphene/phosphorene are considered. 

The examples included in this section are available in the code's GitHub repository in the \texttt{examples} directory \cite{github}.
\subsection{Moir\'e patterns in bilayer graphene (BG)}
\begin{figure}[h]
    \centering
    \includegraphics[width=0.5\textwidth]{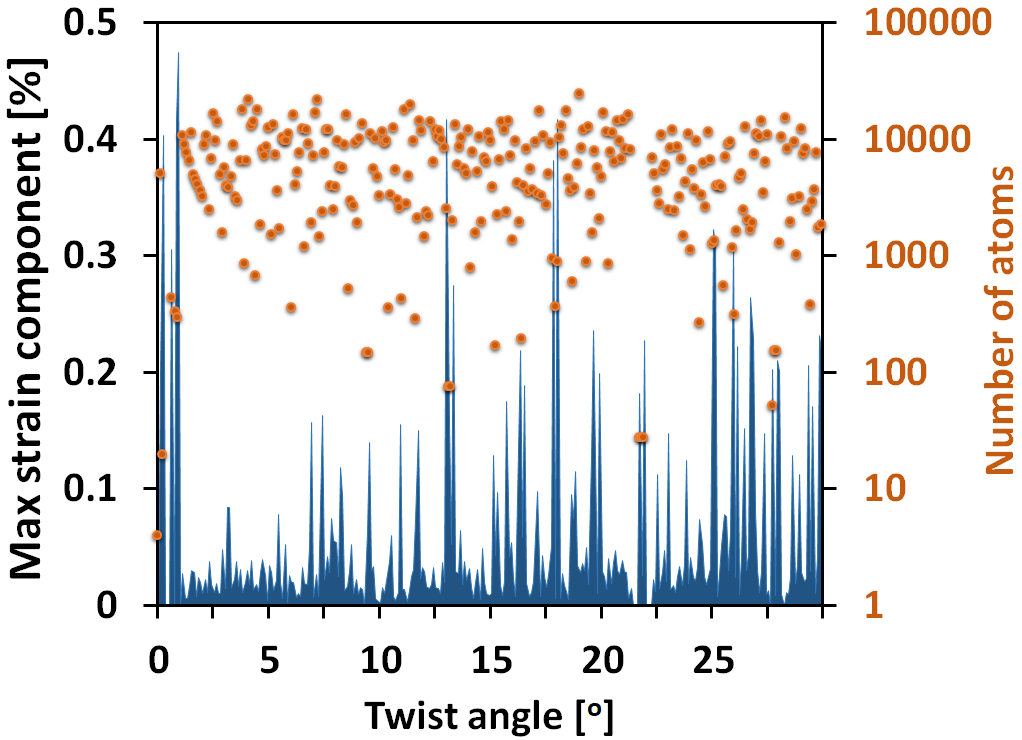}
    \caption{Absolute value of the maximum strain tensor component and number of atoms are presented for the supercells of bilayer graphene as a function of the twist angle. All supercells are generated using the developed software with resolution angle, $\Delta  \theta=0.1^{\circ}$ up to $30^{\circ}$ and an $L_{\max}(A) \leq$ 20 parameter. For clarity, only the results with maximum strain component $\le$ 0.5 $\%$ are presented. Visible dips in the blue bar plot correspond to configurations with moir\'e patterns. Note that there are many twist angles for which the number of atoms is in the hundreds. }
    \label{fig:gr_gr_angle}
\end{figure}

\begin{figure}
    \centering
    \includegraphics[width=0.5\textwidth]{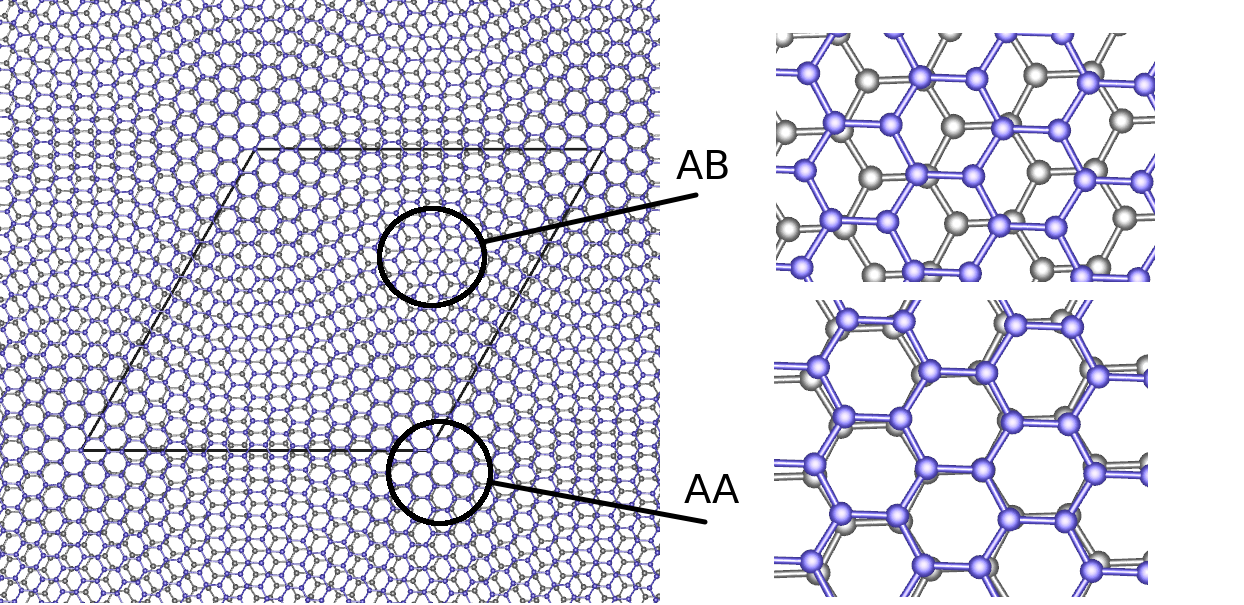}
    \caption{The moir\'e pattern forms on the bilayer graphene with a relative rotation angle (twist angle) between the layers equal to $\theta=6^{\circ}$. A unit cell determined by the software \texttt{supercell-core} is marked in blue and contains 364 atoms. Various local stacking configurations are clearly visible. AA and AB stackings are shown in zoom on the right side of the picture. Grey and violet points indicate the carbon atoms from the bottom and top layers, respectively.}
    \label{fig:moire}
\end{figure}

\begin{table*}[t]
\footnotesize
  \def\arraystretch{1.5}
\caption{Structural information corresponding to bilayer graphene supercells generated by \textit{supercell-core} software. The results are presented for the optimal supercells (small system size and nearly unstrained layers) and for particular twist angles between the graphene layers. The supercells of the bilayer graphene have been generated with twist angle resolution 0.1 $^{\circ}$, and $L_{\max}(A) \leq 20$ parameters. Note, that the supercell vectors can be calculated from $\vec{c_1}=m_{11}\vec{a_1}+m_{21}\vec{a_2}$ and $\vec{c_2}=m_{12}\vec{a_1}+m_{22}\vec{a_2}$ or from 2D rotation matrix and $n_{ij}$ parameters (see Eq. 1). The $m_{ij}$, $n_{ij}$, $\varepsilon_{ij}$ matrices are taken from \textit{log} files. Only maximal strain tensor component is presented.}.
\label{tab:bilayer}
\begin{ruledtabular}
\begin{tabular}{cccccc}
 \begin{tabular}[c]{@{}c@{}}Twist angle\\$\theta [^{\circ}]$ \end{tabular} &
 \begin{tabular}[c]{@{}c@{}}substrate layer (A)\\$M=(m_{ij})$  $\in \mathbb{Z}^{2x2}$ \end{tabular} &
 \begin{tabular}[c]{@{}c@{}}top layer (B)\\$N=(n_{ij})$ $\in \mathbb{Z}^{2x2}$ \end{tabular} &
  \begin{tabular}[c]{@{}c@{}}
  supercell size [\AA$^2$]\\  
  $|\vec{c}_1|\times |\vec{c}_2|\times \sin(\measuredangle(\mathbf{c}_1, \mathbf{c}_2))$
  \end{tabular} &
  \begin{tabular}[c]{@{}c@{}} max. strain component (B$_1$) \\ $\varepsilon_{ji}=(\varepsilon_{ij})^T$ \end{tabular} &
  \begin{tabular}[c]{@{}c@{}}supercell\\ no. atoms\end{tabular}  \\ \hline
1.1 &
  $\begin{pmatrix} 30 & 60 \\ -60 & -29 \end{pmatrix}$ &  
  $\begin{pmatrix}  29 & 60 \\ -60 & -30     \end{pmatrix}$ & $127.8$ \AA $\times127.8$\AA $\times\sin(61.1^{\circ})$
& $\varepsilon_{xx}=\varepsilon_{yy}=0.027\%$  &  10920\\ 
2.0 &
  $\begin{pmatrix}  50 & 33 \\ -49 & -16    \end{pmatrix}$ &
    $\begin{pmatrix} 49 & 33 \\ -50 & -17     \end{pmatrix}$& 
 121.8\AA $\times$ 70.3\AA $\times \sin(30^{\circ})$ 
& $\varepsilon_{xy}=-0.01\%,\varepsilon_{yx}=0.01\%$  &  3268\\ 
  3.9 &
$\begin{pmatrix}  9 & 17 \\ -17 & -8     \end{pmatrix}$ & 
    $\begin{pmatrix}  8 & 17 \\ -17 & -9     \end{pmatrix}$&
 36.2\AA $\times$ 36.2\AA $\times \sin(60^{\circ})$
& $\varepsilon_{xy}=0.019\%,\varepsilon_{yx}=-0.019\%$  &  868\\ 
6.0 &
$\begin{pmatrix} 12 & 37 \\ 21 & 57    \end{pmatrix}$ &
   $\begin{pmatrix}  21 & 62 \\ 12 & 31     \end{pmatrix}$&
 130.7\AA $\times$ 107.5\AA $\times \sin(1.9^{\circ})$ 
& $\varepsilon_{xy}=-0.019\%,\varepsilon_{yx}=0.019\%$  &  364\\
  17.9 &
$\begin{pmatrix} 12 & 37 \\ 21 & 57    \end{pmatrix}$ &
   $\begin{pmatrix}  21 & 62 \\ 12 & 31     \end{pmatrix}$&
 71.2\AA $\times$ 201.7\AA $\times \sin (1.9^{\circ})$ 
& $\varepsilon_{yx}=-0.006\%$  &  372\\ 
  21.8 &
$\begin{pmatrix} 59 & 55 \\ 13 & 12    \end{pmatrix}$ &
    $\begin{pmatrix}  73 & 68 \\ -16 & -15     \end{pmatrix}$&
 163.5\AA $\times$ 152.2\AA $\times \sin(0.1^{\circ})$ 
& $\varepsilon_{yx}=-0.025\%$  &  28\\ 
27.8 &
$\begin{pmatrix} 9 & 11 \\ 42 & 47    \end{pmatrix}$ &
    $\begin{pmatrix}  33 & 38 \\ 21 & 23     \end{pmatrix}$&
 116\AA $\times$ 131.2\AA $\times \sin(0.8^{\circ})$ 
& $\varepsilon_{xy}=-0.008\%$  &  156\\ 
29.4 &
$\begin{pmatrix} 19 & 49 \\ -14 & -31    \end{pmatrix}$ &
    $\begin{pmatrix}  14 & 39 \\ -19 & -46     \end{pmatrix}$&
 42\AA $\times$ 105.6\AA $\times \sin(6.6^{\circ})$
& $\varepsilon_{yx}=-0.028\%$  &  388\\ 
\end{tabular}
\end{ruledtabular}
\end{table*}
Here we examine the construction of the supercell of bilayer graphene, which is a prototypical example of a bilayer system. For this reason, we use a two-atom unit cell of graphene with lattice vectors, $\vec{a}_1=(\frac{\sqrt{3}a}{2}, \frac{-a}{2})$ and $\vec{a}_2=(\frac{\sqrt{3}a}{2}, \frac{a}{2})$, assumed for both layers, where $a = 2.46$ \AA is a lattice constant of graphene. One of the graphene layers can be rotated with respect to the other by a relative angle, $\theta$ (twist angle). Supposing we want to find the twist angles that lead to the appearance of moir\'e patterns, we can use the \texttt{Heterostructure.opt} method from the code, and set the \texttt{algorithm} parameter to \texttt{moire}. It is also recommended to increase the value of the $L_{max}(A)$ (e.g., \texttt{max\_el}  to 20 or 50) in order to catch patterns that are only visible at long distance, and to use a small step for the values of $\theta$ (e.g., 0.025$^{\circ}$). Then, the user can apply the \texttt{log} functionality to assess which angles are most promising (for details see section \ref{description}).

In this example we set \texttt{max$\_$el} to 20 and the range for the twist angle as $\theta \in (0^{\circ}, 30^{\circ})$, with the step equal to 0.1$^{\circ}$. The optimal supercells were found by applying these parameters, and the maximum strain tensor components of these supercells are presented in Fig. \ref{fig:gr_gr_angle}. Particular angles exist when the supercells are small (tens to hundreds of atoms) and have low strain values ($<0.03\%$). Some of the lowest-strain angles are collected in Table \ref{tab:bilayer}. In the case where two identical layers with a twist angle equal to $0^{\circ}$ are considered, the strain tensor is exactly zero, and the generated supercell is identical to the primary cell. Because of this obvious result, we do not include this case in Table \ref{tab:bilayer}. Note that the lattice mismatch is commonly calculated as the difference between the lattice constants, even for rotated layers with strain distributions
that are different from those determined by this simple approximation. Therefore, supercells resulting from construction of such twisted layers are assumed to exhibit low strain values, although, in fact, the strain components (e.g., shear strains) can be substantial, and can influence the material's electronic properties. The number of atoms in the substrate layer can be calculated using the equation, $det(M)\,\times  N_{at}$, where $N_{at}$ is the number of atoms in primary cell, and M is equal to $M=(m_{ij})$. The number of atoms in other layers can be calculated in an analogous manner.

The results (Table \ref{tab:bilayer}) demonstrate that, for small twist angles, the size of the supercell is large and the number of atoms is high, whereas for large twist angles, the situation is reversed. These results confirm a well-known fact observed for twisted bilayer structures where the relevant length scale is on the order of $1/\theta$ \cite{TramblydeLaissardire2010}. In particular, for small twist angles such as $1.1^{\circ}$ or $2^{\circ}$, the periodicity becomes large enough that it cannot be handled as lattice inputs for DFT codes. However, other potential approaches, such as tight binding methods (TB) \cite{PythonTB}, can be used instead. Moreover, our \texttt{supercell-core} software successfully identifies the angles that have been reported previously based on STM experiments, specifically the $21.8\%$ \cite{PhysRevB.78.125406} and $1.1\%$ \cite{Cao2018} for bilayer systems. In addition, we present the visualisation of the optimal supercell generated for the twist angle of $6^{\circ}$ as shown in Fig. 4, where the moir\'e pattern is clearly visible along with various stacking configurations. The different stacking configurations of the bilayer graphene and theirs crucial impact on the electronic properties have been highlighted in many papers so far (see i.e. Ref.\cite{APP2011.Birowska}), however, for a twisted  layers many stacking configurations are automatically included within one supercell (see e.g. Fig. \ref{fig:moire}). 

\begin{table*}[t!]
\footnotesize
  \def\arraystretch{1.5}
\caption{Structural information corresponding to the optimal supercells generated by \textit{supercell-core} software for the trilayer systems: trilayer graphene (TG) and hexagonal boron nitride/graphene/phosphorene (hBN/G/P). The supercells of the trilayer graphene have been generated with twist angle resolution equl to 0.1 $^{\circ}$, and $L_{\max}(A) \leq 80$. 
Only maximal strain tensor component is presented. The twist angles $\theta_1$, $\theta_2$ correspond to relative rotation of the graphene and phosphorene layers in respect to the hBN layer, respectively. For the clarity of presentation, only one of the maximal strain components is presented, however, in some of the cases presented here the $|\varepsilon_{xy}|=|\varepsilon_{yx}|$ or $|\varepsilon_{xx}|=|\varepsilon_{yy}|$.}
\label{tab:hbn}
\begin{ruledtabular}
\begin{tabular}{cccccccc}
\begin{tabular}[c]{@{}c@{}} System\\$L_{max}(A)$ \end{tabular}  &\begin{tabular}[c]{@{}c@{}}Twist angles\\$\theta_1 [^{\circ}], \theta_2 [^{\circ}]$ \end{tabular} &
 \begin{tabular}[c]{@{}c@{}}bottom \\ layer (A)\\$M=(m_{ij})$\\  $\in \mathbb{Z}^{2x2}$ \end{tabular} &
  \begin{tabular}[c]{@{}c@{}}middle \\ layer (B$_1$)\\$N=(n_{ij})$\\  $\in \mathbb{Z}^{2x2}$ \end{tabular} &
 \begin{tabular}[c]{@{}c@{}}top layer (B$_2$)\\$N'=(n'_{ij})$\\  $\in \mathbb{Z}^{2x2}$ \end{tabular} &
\begin{tabular}[c]{@{}c@{}}
  supercell size [$\AA^2$]\\  
  $|\vec{c}_1|\times |\vec{c}_2|\times \sin(\measuredangle(\mathbf{c}_1, \mathbf{c}_2))$
  \end{tabular}  &
  \begin{tabular}[c]{@{}l@{}}max. strain component \\ layer B$_1$; layer B$_2$\\ ($\%$)\end{tabular} &
  \begin{tabular}[c]{@{}l@{}}
  No. \\ atoms
  \end{tabular}  \\ \hline
TG &$0, 21.8$ &
$\begin{pmatrix}  12 & 19 \\ 3 & 3     \end{pmatrix}$ & 
$\begin{pmatrix}  12 & 19 \\ 3 & 3     \end{pmatrix}$ &
$\begin{pmatrix}  15 & 23 \\ -3 & -6     \end{pmatrix}$ &
 $33.8$\AA $\times  50.8$\AA $\times Sin(3.7^{\circ})$&
    0$\%$; $\varepsilon_{yx}=-0.005\%$& 126  \\ 
    TG &$21.8, 17.9$ &
$\begin{pmatrix}  19 & 16 \\ 10 & -3     \end{pmatrix}$ & 
$\begin{pmatrix}  26 & 17 \\ -1 & -9     \end{pmatrix}$ &
$\begin{pmatrix}  25 & 17 \\ 1 & -8     \end{pmatrix}$ &
 62.8\AA $\times$  36.2\AA $\times Sin(30^{\circ})$&
    $\varepsilon_{yx}=-0.03\%$; $\varepsilon_{yx}=-0.005\%$& 1302  \\ 
TG &$6.0, 27.9$ &
$\begin{pmatrix}  17 & 11 \\ -16 & -5     \end{pmatrix}$ & 
$\begin{pmatrix}  16 & 11 \\ -17 & -6     \end{pmatrix}$ &
$\begin{pmatrix}  11 & 10 \\ -19 & -9     \end{pmatrix}$ &
 40.6\AA $\times$  23.5\AA $\times Sin(30^{\circ})$&
    $\varepsilon_{xy}=-0.019\%$; $\varepsilon_{yx}=-0.2\%$& 546  \\ 
\begin{tabular}[c]{@{}c@{}}hBN/G/P\\$L_{max}(A)\leq10$ \end{tabular} &$6.7, 0.8$ &
$\begin{pmatrix}  4 & 6 \\ -10 & 3     \end{pmatrix}$ & 
$\begin{pmatrix}  3 & 7 \\ -10 & 2     \end{pmatrix}$ &
$\begin{pmatrix}  -4 & 6 \\  -4& -1 \end{pmatrix}$ &
  21.8\AA $\times$ 19.9\AA $\times Sin(115.7^{\circ})$    &
  $\varepsilon_{xx}$=1.7$\%$; $\varepsilon_{yx}$=-1.9$\%$ & 403 \\
\begin{tabular}[c]{@{}c@{}}hBN/G/P\\$L_{max}(A)\leq10$ \end{tabular} &$10.9, 29.9$ &
$\begin{pmatrix}  6 & 6 \\ -3 & 6     \end{pmatrix}$ & 
$\begin{pmatrix}  6 & 8 \\ -4 & 4     \end{pmatrix}$ &
$\begin{pmatrix}  0 & 7 \\  -3& -3 \end{pmatrix}$ &
  13\AA $\times$ 26\AA $\times Sin(60^{\circ})$    &
 $\varepsilon_{xx}$=0.11$\%$; $\varepsilon_{xy}$=-1.31$\%$ & 302 \\ 
\begin{tabular}[c]{@{}c@{}}hBN/G/P\\$L_{max}(A)\leq10$ \end{tabular} &$11.5, 29.8$ &
$\begin{pmatrix}  8 & 6 \\ 9 & -3     \end{pmatrix}$ & 
$\begin{pmatrix}  11 & 6 \\ 6 & -4     \end{pmatrix}$ &
$\begin{pmatrix}  10 & 0 \\  -4& -3 \end{pmatrix}$ &
  36.9\AA $\times$ 13 \AA $\times Sin(62^{\circ})$    &
  $\varepsilon_{yy}$=-0.92$\%$;  $\varepsilon_{yx}$=0.23$\%$ & 437 \\ 
\begin{tabular}[c]{@{}c@{}}hBN/G/P\\$L_{max}(A)\leq40$ \end{tabular}&$23.1, 4.4$ &
$\begin{pmatrix}  13 & 17 \\ 17 & -30     \end{pmatrix}$ & 
$\begin{pmatrix}  23 & 6 \\ 6 & -29     \end{pmatrix}$ &
$\begin{pmatrix}  20 & -10 \\ 0 & -13     \end{pmatrix}$ &
 65.2\AA $\times$ 65.2 \AA $\times Sin(120^{\circ})$&
   $\varepsilon_{xx}$=0.03$\%$;  $\varepsilon_{yy}$=0.09$\%$& 3802 \\ 
\end{tabular}
\end{ruledtabular}
\end{table*}

\subsection{Trilayer systems: trilayer graphene (TG) and hBN/graphene/phosphorene}
The natural extension of the bilayer system is to simply add another layer. In principle, our code allows the user to find the optimal supercell for $n$ layers exhibiting different types of symmetry.

Thus, in this example, we consider a trilayer van der Waals heterostructure consisting of hexagonal boron nitride, graphene, and phosphorene. This structure is significantly more complicated than the previous example. Although graphene and hBN share the same hexagonal symmetry and have similar lattice constants ($2.46$ \AA and $2.52$\AA, respectively), phosphorene  has a rectangular lateral cell with lattice constants equal to $a=3.26$ \AA and $b=4.35$ \AA (optimised lattice constants obtained in LDA approximation for monolayer taken from \cite{Birowska_2019}, (see Fig. \ref{fig:hbn_lattices}). These differences between the layers makes finding an optimal supercell more difficult. 

\begin{figure}
    \centering
    \includegraphics[width=0.46\textwidth]{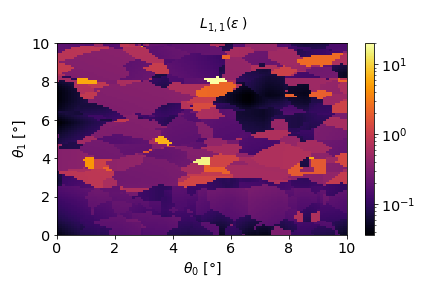}
    \caption{Strain distribution for an hBN/graphene/phosphorene heterostructure as a function of the twist angles, $\theta_0$ and $\theta_1$. For the clarity of the presentation, each angle is presented within the range ($0^{\circ},10^{\circ}$) and step size of $0.1^{\circ}$. The black colour indicates for the range of angles with the least amount of strain.}.
    
    \label{fig:hbn_map}
\end{figure}

\begin{figure}
    \centering
    \includegraphics[width=0.5\textwidth]{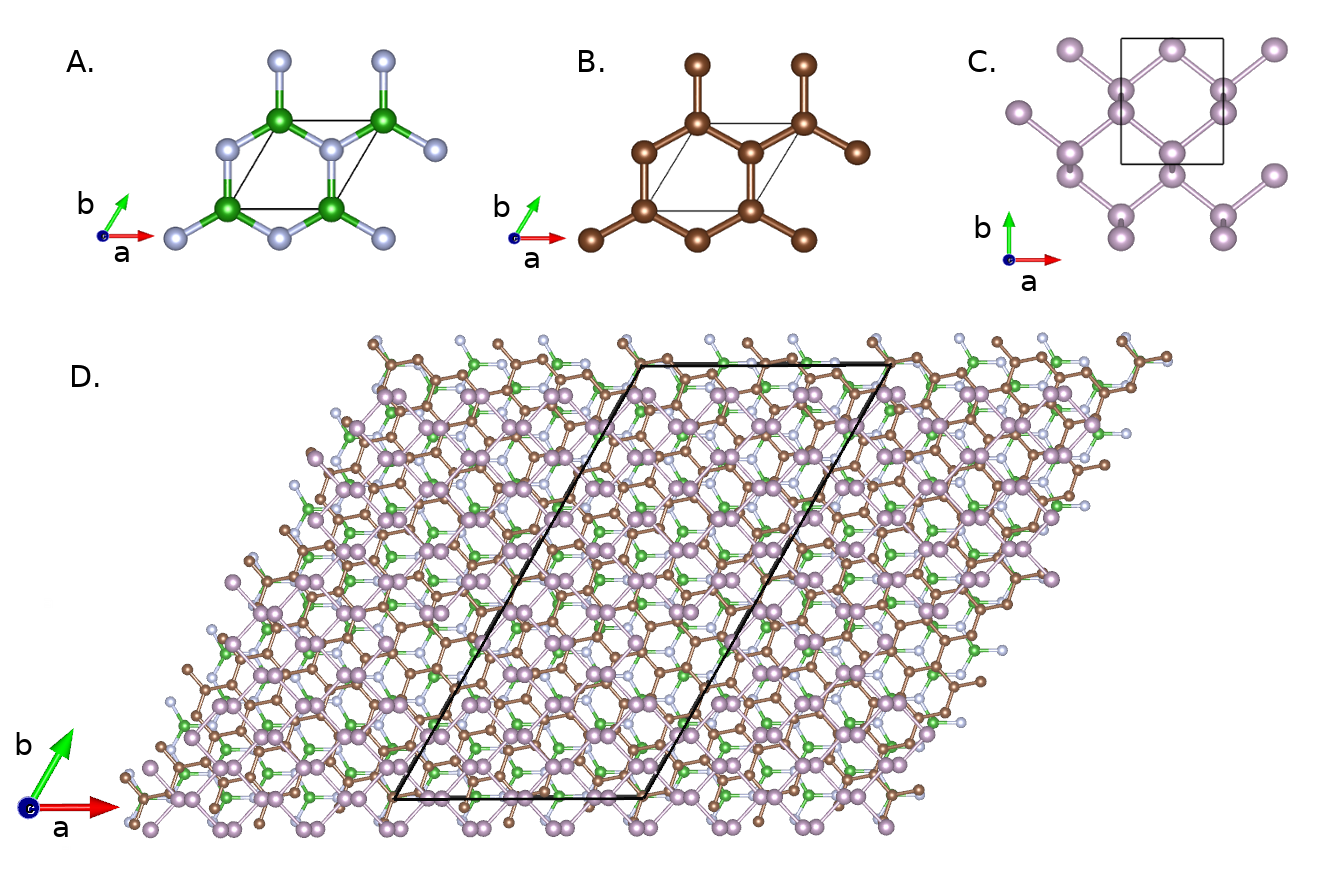}
    \caption{The primary cells of the A. hBN, B. graphene, and C. phoshorene layers used for constructing the optimal van der Waals trilayer hBN/G/P supercell and D. an example of an optimal supercell for a twisted trilayer hBN/G/P structure with twist angles equal to $\theta_1=10.9^{\circ}$ and $\theta_2=29.9^{\circ}$, corresponding to graphene and phosphorene rotations with respect to the hBN layer.}
    \label{fig:hbn_lattices}
\end{figure}

The \texttt{supercell-core} software can be used to find the optimal configuration for this example with any combination of twist angles. 
The layer strains resulting from the generated supercells are presented in Fig. \ref{fig:hbn_map} as a function of two twist angles, $\theta_1$ and $\theta_2$. These angles ($\theta_1$, $\theta_2$) correspond to the relative rotation of the graphene (layer $B_1$) and phosporene (layer $B_2$) layers with respect to the hBN layer (layer A), respectively. The dark navy blue regions in Fig. \ref{fig:hbn_map} indicate the set of angles that correspond to supercells with relatively low strains. Table \ref{tab:hbn} provides the details regarding a few generated supercells that contain hundreds of atoms. The results clearly show that the strains in these supercells are relatively high.  This is because we constrained the supercell size ($L_{max}(A)\le 10$) so that the number of atoms is less than one thousand, while our system is composed of three incommensurate lattices. This is a fundamental problem of the system, and not a problem of the program. To find better supercells, one can loosen the constraints on the supercell size, although this introduces the risk of finding supercells that are too big for DFT calculations. In order to minimise the strain, it is recommended that the user increases the value of the  $L_{max}(A)$ parameter. By increasing this parameter, one directly enlarges the supercell size considered, and thus, minimises the strain (see the last raw results in Table \ref{tab:hbn}). Increasing the resolution of the twist angle may also help, in principle, but there is little chance that an optimal supercell would not be found with a resolution on the order of $0.1^{\circ}$.

In the Table \ref{tab:hbn}, we also present  selected  supercells  for trilayer graphene (TG). Note, that whenever one of the twist angles is zero, the results correspond to bilayer system (first raw in Table \ref{tab:hbn} for $\theta_1=0^{\circ}$, $\theta_2=21.8^{\circ}$). However, the software found different result than presented for bilayer system (see Table \ref{tab:bilayer} for $\theta_1=21.8^{\circ}$). When \texttt{supercell-core} finds a true moir\'e pattern, any parallelogram with vertices on the repeated patterns' nodes is a zero-strain supercell.  The program should then find the smallest of such supercells. However, the calculated strain for each of those supercell is non-zero because of floating point errors. When choosing the optimal supercell, the program has some fixed tolerancy (epsilon) of the variation in the strain value but sometimes it happens to be too low to find the smallest supercell from a set of equally good supercells. A potential solution is to re-run the calculations for low-strain supercells with lower value of the $L_{max}(A)$ parameter.


Generally, for cases where $n>3$, it is difficult to obtain the commensurability condition for a relatively small system size, especially one with different types of layers that exhibit different symmetries.

{}

{}

\section{CONCLUSIONS}

 In this report, we present the novel \texttt{supercell-core} software  which has been developed to facilitate construction and analysis of vertically-aligned 2D layers. In principle, the code is designed to handle structures comprised of "n"  different types of layers exhibiting different lateral symmetries. The developed software allows the user to find the optimal supercell, with the smallest number of atoms and lowest strains experienced by adjacent layers. The methodology is based on a commensurability condition, which implies long-range crystalline order in the sets of lattice planes that make up the van der Waals heterostructure. This condition is enforced by applying strain to the top layers of the structures. In the described approach, the bottom layer is always unstrained.
 
The software works with POSCAR files, and it is therefore compatible with the DFT software, VASP, and with the widely-used visualisation software, VESTA. In addition, there are two algorithms that are implemented, namely \texttt{Direct} and \texttt{Fast}. The former spans all possible configurations from which the supercells can be constructed, while the latter is more efficient and designed to search for unstrained configurations (e.g., those exhibiting moir\'e patterns). The former is more hefty and resource-intensive, while the latter is a more efficient algorithm.

Depending on the user's requirements, the developed software also enables construction of the optimal supercells  based on the twist angle(s) between the layers. It also allows users to study strained supercells in order to examine the impact of the strain distribution on the various properties of van der Waals heterostructures. Both types of results can be used as structural inputs for further calculations based on an \textit{ab initio} approach (using DFT software, such as VASP, Quantum Espresso, or SIESTA). Particularly for cases of large supercells containing thousands of atoms, the results provided by our software can be used as structural inputs for software based on tight binding methods.

\section*{Acknowledgements}
This work is funded by the National Science Centre, Poland, grant no. UMO-2016/23/D/ST3/03446. Access to the computing facilities of PL-Grid Polish  
Infrastructure for Supporting Computational Science in the European Research Space, and acknowledge access to the computing facilities of the Interdisciplinary Centre of Modeling (ICM), University of Warsaw. We made use of computing resources of TU Dresden ZIH within the project \textit{TransPheMat}.

\bibliography{supercell}

\providecommand{\noopsort}[1]{}\providecommand{\singleletter}[1]{#1}%
\begin{thebibliography}{33}%
\makeatletter
\providecommand \@ifxundefined [1]{%
 \@ifx{#1\undefined}
}%
\providecommand \@ifnum [1]{%
 \ifnum #1\expandafter \@firstoftwo
 \else \expandafter \@secondoftwo
 \fi
}%
\providecommand \@ifx [1]{%
 \ifx #1\expandafter \@firstoftwo
 \else \expandafter \@secondoftwo
 \fi
}%
\providecommand \natexlab [1]{#1}%
\providecommand \enquote  [1]{``#1''}%
\providecommand \bibnamefont  [1]{#1}%
\providecommand \bibfnamefont [1]{#1}%
\providecommand \citenamefont [1]{#1}%
\providecommand \href@noop [0]{\@secondoftwo}%
\providecommand \href [0]{\begingroup \@sanitize@url \@href}%
\providecommand \@href[1]{\@@startlink{#1}\@@href}%
\providecommand \@@href[1]{\endgroup#1\@@endlink}%
\providecommand \@sanitize@url [0]{\catcode `\\12\catcode `\$12\catcode
  `\&12\catcode `\#12\catcode `\^12\catcode `\_12\catcode `\%12\relax}%
\providecommand \@@startlink[1]{}%
\providecommand \@@endlink[0]{}%
\providecommand \url  [0]{\begingroup\@sanitize@url \@url }%
\providecommand \@url [1]{\endgroup\@href {#1}{\urlprefix }}%
\providecommand \urlprefix  [0]{URL }%
\providecommand \Eprint [0]{\href }%
\providecommand \doibase [0]{http://dx.doi.org/}%
\providecommand \selectlanguage [0]{\@gobble}%
\providecommand \bibinfo  [0]{\@secondoftwo}%
\providecommand \bibfield  [0]{\@secondoftwo}%
\providecommand \translation [1]{[#1]}%
\providecommand \BibitemOpen [0]{}%
\providecommand \bibitemStop [0]{}%
\providecommand \bibitemNoStop [0]{.\EOS\space}%
\providecommand \EOS [0]{\spacefactor3000\relax}%
\providecommand \BibitemShut  [1]{\csname bibitem#1\endcsname}%
\let\auto@bib@innerbib\@empty
\bibitem [{\citenamefont {Cisternas}, \citenamefont {Flores},\ and\
  \citenamefont {Vargas}(2008)}]{PhysRevB.78.125406}%
  \BibitemOpen
  \bibfield  {author} {\bibinfo {author} {\bibfnamefont {E.}~\bibnamefont
  {Cisternas}}, \bibinfo {author} {\bibfnamefont {M.}~\bibnamefont {Flores}}, \
  and\ \bibinfo {author} {\bibfnamefont {P.}~\bibnamefont {Vargas}},\
  }\bibfield  {title} {\enquote {\bibinfo {title} {Superstructures in arrays of
  rotated graphene layers: Electronic structure calculations},}\ }\href
  {\doibase 10.1103/PhysRevB.78.125406} {\bibfield  {journal} {\bibinfo
  {journal} {Phys. Rev. B}\ }\textbf {\bibinfo {volume} {78}},\ \bibinfo
  {pages} {125406} (\bibinfo {year} {2008})}\BibitemShut {NoStop}%
\bibitem [{\citenamefont {Ohta}\ \emph {et~al.}(2012)\citenamefont {Ohta},
  \citenamefont {Robinson}, \citenamefont {Feibelman}, \citenamefont
  {Bostwick}, \citenamefont {Rotenberg},\ and\ \citenamefont
  {Beechem}}]{PhysRevLett.109.186807}%
  \BibitemOpen
  \bibfield  {author} {\bibinfo {author} {\bibfnamefont {T.}~\bibnamefont
  {Ohta}}, \bibinfo {author} {\bibfnamefont {J.~T.}\ \bibnamefont {Robinson}},
  \bibinfo {author} {\bibfnamefont {P.~J.}\ \bibnamefont {Feibelman}}, \bibinfo
  {author} {\bibfnamefont {A.}~\bibnamefont {Bostwick}}, \bibinfo {author}
  {\bibfnamefont {E.}~\bibnamefont {Rotenberg}}, \ and\ \bibinfo {author}
  {\bibfnamefont {T.~E.}\ \bibnamefont {Beechem}},\ }\bibfield  {title}
  {\enquote {\bibinfo {title} {Evidence for interlayer coupling and moir\'e
  periodic potentials in twisted bilayer graphene},}\ }\href {\doibase
  10.1103/PhysRevLett.109.186807} {\bibfield  {journal} {\bibinfo  {journal}
  {Phys. Rev. Lett.}\ }\textbf {\bibinfo {volume} {109}},\ \bibinfo {pages}
  {186807} (\bibinfo {year} {2012})}\BibitemShut {NoStop}%
\bibitem [{\citenamefont {Hunt}\ \emph {et~al.}(2013)\citenamefont {Hunt},
  \citenamefont {Sanchez-Yamagishi}, \citenamefont {Young}, \citenamefont
  {Yankowitz}, \citenamefont {LeRoy}, \citenamefont {Watanabe}, \citenamefont
  {Taniguchi}, \citenamefont {Moon}, \citenamefont {Koshino}, \citenamefont
  {Jarillo-Herrero},\ and\ \citenamefont {Ashoori}}]{Hunt1427}%
  \BibitemOpen
  \bibfield  {author} {\bibinfo {author} {\bibfnamefont {B.}~\bibnamefont
  {Hunt}}, \bibinfo {author} {\bibfnamefont {J.~D.}\ \bibnamefont
  {Sanchez-Yamagishi}}, \bibinfo {author} {\bibfnamefont {A.~F.}\ \bibnamefont
  {Young}}, \bibinfo {author} {\bibfnamefont {M.}~\bibnamefont {Yankowitz}},
  \bibinfo {author} {\bibfnamefont {B.~J.}\ \bibnamefont {LeRoy}}, \bibinfo
  {author} {\bibfnamefont {K.}~\bibnamefont {Watanabe}}, \bibinfo {author}
  {\bibfnamefont {T.}~\bibnamefont {Taniguchi}}, \bibinfo {author}
  {\bibfnamefont {P.}~\bibnamefont {Moon}}, \bibinfo {author} {\bibfnamefont
  {M.}~\bibnamefont {Koshino}}, \bibinfo {author} {\bibfnamefont
  {P.}~\bibnamefont {Jarillo-Herrero}}, \ and\ \bibinfo {author} {\bibfnamefont
  {R.~C.}\ \bibnamefont {Ashoori}},\ }\bibfield  {title} {\enquote {\bibinfo
  {title} {Massive dirac fermions and hofstadter butterfly in a van der waals
  heterostructure},}\ }\href {\doibase 10.1126/science.1237240} {\bibfield
  {journal} {\bibinfo  {journal} {Science}\ }\textbf {\bibinfo {volume}
  {340}},\ \bibinfo {pages} {1427--1430} (\bibinfo {year} {2013})},\ \Eprint
  {http://arxiv.org/abs/https://science.sciencemag.org/content/340/6139/1427.full.pdf}
  {https://science.sciencemag.org/content/340/6139/1427.full.pdf} \BibitemShut
  {NoStop}%
\bibitem [{\citenamefont {Cao}\ \emph {et~al.}(2018{\natexlab{a}})\citenamefont
  {Cao}, \citenamefont {Fatemi}, \citenamefont {Demir}, \citenamefont {Fang},
  \citenamefont {Tomarken}, \citenamefont {Luo}, \citenamefont
  {Sanchez-Yamagishi}, \citenamefont {Watanabe}, \citenamefont {Taniguchi},
  \citenamefont {Kaxiras}, \citenamefont {Ashoori},\ and\ \citenamefont
  {Jarillo-Herrero}}]{Cao2018ins}%
  \BibitemOpen
  \bibfield  {author} {\bibinfo {author} {\bibfnamefont {Y.}~\bibnamefont
  {Cao}}, \bibinfo {author} {\bibfnamefont {V.}~\bibnamefont {Fatemi}},
  \bibinfo {author} {\bibfnamefont {A.}~\bibnamefont {Demir}}, \bibinfo
  {author} {\bibfnamefont {S.}~\bibnamefont {Fang}}, \bibinfo {author}
  {\bibfnamefont {S.~L.}\ \bibnamefont {Tomarken}}, \bibinfo {author}
  {\bibfnamefont {J.~Y.}\ \bibnamefont {Luo}}, \bibinfo {author} {\bibfnamefont
  {J.~D.}\ \bibnamefont {Sanchez-Yamagishi}}, \bibinfo {author} {\bibfnamefont
  {K.}~\bibnamefont {Watanabe}}, \bibinfo {author} {\bibfnamefont
  {T.}~\bibnamefont {Taniguchi}}, \bibinfo {author} {\bibfnamefont
  {E.}~\bibnamefont {Kaxiras}}, \bibinfo {author} {\bibfnamefont {R.~C.}\
  \bibnamefont {Ashoori}}, \ and\ \bibinfo {author} {\bibfnamefont
  {P.}~\bibnamefont {Jarillo-Herrero}},\ }\bibfield  {title} {\enquote
  {\bibinfo {title} {Correlated insulator behaviour at half-filling in
  magic-angle graphene superlattices},}\ }\href {\doibase 10.1038/nature26154}
  {\bibfield  {journal} {\bibinfo  {journal} {Nature}\ }\textbf {\bibinfo
  {volume} {556}},\ \bibinfo {pages} {80--84} (\bibinfo {year}
  {2018}{\natexlab{a}})}\BibitemShut {NoStop}%
\bibitem [{\citenamefont {Zhu}\ \emph {et~al.}(2020)\citenamefont {Zhu},
  \citenamefont {Cazeaux}, \citenamefont {Luskin},\ and\ \citenamefont
  {Kaxiras}}]{PhysRevB.101.224107}%
  \BibitemOpen
  \bibfield  {author} {\bibinfo {author} {\bibfnamefont {Z.}~\bibnamefont
  {Zhu}}, \bibinfo {author} {\bibfnamefont {P.}~\bibnamefont {Cazeaux}},
  \bibinfo {author} {\bibfnamefont {M.}~\bibnamefont {Luskin}}, \ and\ \bibinfo
  {author} {\bibfnamefont {E.}~\bibnamefont {Kaxiras}},\ }\bibfield  {title}
  {\enquote {\bibinfo {title} {Modeling mechanical relaxation in incommensurate
  trilayer van der waals heterostructures},}\ }\href {\doibase
  10.1103/PhysRevB.101.224107} {\bibfield  {journal} {\bibinfo  {journal}
  {Phys. Rev. B}\ }\textbf {\bibinfo {volume} {101}},\ \bibinfo {pages}
  {224107} (\bibinfo {year} {2020})}\BibitemShut {NoStop}%
\bibitem [{\citenamefont {Kang}\ \emph {et~al.}(2017)\citenamefont {Kang},
  \citenamefont {Zhang}, \citenamefont {Michaud-Rioux}, \citenamefont {Kong},
  \citenamefont {Hu}, \citenamefont {Yu},\ and\ \citenamefont
  {Guo}}]{PhysRevB.96.195406}%
  \BibitemOpen
  \bibfield  {author} {\bibinfo {author} {\bibfnamefont {P.}~\bibnamefont
  {Kang}}, \bibinfo {author} {\bibfnamefont {W.-T.}\ \bibnamefont {Zhang}},
  \bibinfo {author} {\bibfnamefont {V.}~\bibnamefont {Michaud-Rioux}}, \bibinfo
  {author} {\bibfnamefont {X.-H.}\ \bibnamefont {Kong}}, \bibinfo {author}
  {\bibfnamefont {C.}~\bibnamefont {Hu}}, \bibinfo {author} {\bibfnamefont
  {G.-H.}\ \bibnamefont {Yu}}, \ and\ \bibinfo {author} {\bibfnamefont
  {H.}~\bibnamefont {Guo}},\ }\bibfield  {title} {\enquote {\bibinfo {title}
  {Moir\'e impurities in twisted bilayer black phosphorus: Effects on the
  carrier mobility},}\ }\href {\doibase 10.1103/PhysRevB.96.195406} {\bibfield
  {journal} {\bibinfo  {journal} {Phys. Rev. B}\ }\textbf {\bibinfo {volume}
  {96}},\ \bibinfo {pages} {195406} (\bibinfo {year} {2017})}\BibitemShut
  {NoStop}%
\bibitem [{\citenamefont {Decker}\ \emph {et~al.}(2011)\citenamefont {Decker},
  \citenamefont {Wang}, \citenamefont {Brar}, \citenamefont {Regan},
  \citenamefont {Tsai}, \citenamefont {Wu}, \citenamefont {Gannett},
  \citenamefont {Zettl},\ and\ \citenamefont {Crommie}}]{Decker2011}%
  \BibitemOpen
  \bibfield  {author} {\bibinfo {author} {\bibfnamefont {R.}~\bibnamefont
  {Decker}}, \bibinfo {author} {\bibfnamefont {Y.}~\bibnamefont {Wang}},
  \bibinfo {author} {\bibfnamefont {V.~W.}\ \bibnamefont {Brar}}, \bibinfo
  {author} {\bibfnamefont {W.}~\bibnamefont {Regan}}, \bibinfo {author}
  {\bibfnamefont {H.-Z.}\ \bibnamefont {Tsai}}, \bibinfo {author}
  {\bibfnamefont {Q.}~\bibnamefont {Wu}}, \bibinfo {author} {\bibfnamefont
  {W.}~\bibnamefont {Gannett}}, \bibinfo {author} {\bibfnamefont
  {A.}~\bibnamefont {Zettl}}, \ and\ \bibinfo {author} {\bibfnamefont {M.~F.}\
  \bibnamefont {Crommie}},\ }\bibfield  {title} {\enquote {\bibinfo {title}
  {Local electronic properties of graphene on a bn substrate via scanning
  tunneling microscopy},}\ }\href {\doibase 10.1021/nl2005115} {\bibfield
  {journal} {\bibinfo  {journal} {Nano Letters}\ }\textbf {\bibinfo {volume}
  {11}},\ \bibinfo {pages} {2291--2295} (\bibinfo {year} {2011})}\BibitemShut
  {NoStop}%
\bibitem [{\citenamefont {Woods}\ \emph {et~al.}(2014)\citenamefont {Woods},
  \citenamefont {Britnell}, \citenamefont {Eckmann}, \citenamefont {Ma},
  \citenamefont {Lu}, \citenamefont {Guo}, \citenamefont {Lin}, \citenamefont
  {Yu}, \citenamefont {Cao}, \citenamefont {Gorbachev}, \citenamefont
  {Kretinin}, \citenamefont {Park}, \citenamefont {Ponomarenko}, \citenamefont
  {Katsnelson}, \citenamefont {Gornostyrev}, \citenamefont {Watanabe},
  \citenamefont {Taniguchi}, \citenamefont {Casiraghi}, \citenamefont {Gao},
  \citenamefont {Geim},\ and\ \citenamefont {Novoselov}}]{Woods2014}%
  \BibitemOpen
  \bibfield  {author} {\bibinfo {author} {\bibfnamefont {C.~R.}\ \bibnamefont
  {Woods}}, \bibinfo {author} {\bibfnamefont {L.}~\bibnamefont {Britnell}},
  \bibinfo {author} {\bibfnamefont {A.}~\bibnamefont {Eckmann}}, \bibinfo
  {author} {\bibfnamefont {R.~S.}\ \bibnamefont {Ma}}, \bibinfo {author}
  {\bibfnamefont {J.~C.}\ \bibnamefont {Lu}}, \bibinfo {author} {\bibfnamefont
  {H.~M.}\ \bibnamefont {Guo}}, \bibinfo {author} {\bibfnamefont
  {X.}~\bibnamefont {Lin}}, \bibinfo {author} {\bibfnamefont {G.~L.}\
  \bibnamefont {Yu}}, \bibinfo {author} {\bibfnamefont {Y.}~\bibnamefont
  {Cao}}, \bibinfo {author} {\bibfnamefont {R.~.~V.}\ \bibnamefont
  {Gorbachev}}, \bibinfo {author} {\bibfnamefont {A.~V.}\ \bibnamefont
  {Kretinin}}, \bibinfo {author} {\bibfnamefont {J.}~\bibnamefont {Park}},
  \bibinfo {author} {\bibfnamefont {L.~A.}\ \bibnamefont {Ponomarenko}},
  \bibinfo {author} {\bibfnamefont {M.~I.}\ \bibnamefont {Katsnelson}},
  \bibinfo {author} {\bibfnamefont {Y.~N.}\ \bibnamefont {Gornostyrev}},
  \bibinfo {author} {\bibfnamefont {K.}~\bibnamefont {Watanabe}}, \bibinfo
  {author} {\bibfnamefont {T.}~\bibnamefont {Taniguchi}}, \bibinfo {author}
  {\bibfnamefont {C.}~\bibnamefont {Casiraghi}}, \bibinfo {author}
  {\bibfnamefont {H.-J.}\ \bibnamefont {Gao}}, \bibinfo {author} {\bibfnamefont
  {A.~K.}\ \bibnamefont {Geim}}, \ and\ \bibinfo {author} {\bibfnamefont
  {K.~.~S.}\ \bibnamefont {Novoselov}},\ }\bibfield  {title} {\enquote
  {\bibinfo {title} {Commensurate-incommensurate transition in graphene on
  hexagonal boron nitride},}\ }\href {\doibase 10.1038/nphys2954} {\bibfield
  {journal} {\bibinfo  {journal} {Nature Physics}\ }\textbf {\bibinfo {volume}
  {10}},\ \bibinfo {pages} {451--456} (\bibinfo {year} {2014})}\BibitemShut
  {NoStop}%
\bibitem [{\citenamefont {Kang}\ \emph {et~al.}(2013)\citenamefont {Kang},
  \citenamefont {Li}, \citenamefont {Li}, \citenamefont {Xia},\ and\
  \citenamefont {Wang}}]{Kang2013}%
  \BibitemOpen
  \bibfield  {author} {\bibinfo {author} {\bibfnamefont {J.}~\bibnamefont
  {Kang}}, \bibinfo {author} {\bibfnamefont {J.}~\bibnamefont {Li}}, \bibinfo
  {author} {\bibfnamefont {S.-S.}\ \bibnamefont {Li}}, \bibinfo {author}
  {\bibfnamefont {J.-B.}\ \bibnamefont {Xia}}, \ and\ \bibinfo {author}
  {\bibfnamefont {L.-W.}\ \bibnamefont {Wang}},\ }\bibfield  {title} {\enquote
  {\bibinfo {title} {Electronic structural moir{\'e} pattern effects on
  mos2/mose2 2d heterostructures},}\ }\href {\doibase 10.1021/nl4030648}
  {\bibfield  {journal} {\bibinfo  {journal} {Nano Letters}\ }\textbf {\bibinfo
  {volume} {13}},\ \bibinfo {pages} {5485--5490} (\bibinfo {year}
  {2013})}\BibitemShut {NoStop}%
\bibitem [{\citenamefont {Alexeev}\ \emph {et~al.}(2019)\citenamefont
  {Alexeev}, \citenamefont {Ruiz-Tijerina}, \citenamefont {Danovich},
  \citenamefont {Hamer}, \citenamefont {Terry}, \citenamefont {Nayak},
  \citenamefont {Ahn}, \citenamefont {Pak}, \citenamefont {Lee}, \citenamefont
  {Sohn}, \citenamefont {Molas}, \citenamefont {Koperski}, \citenamefont
  {Watanabe}, \citenamefont {Taniguchi}, \citenamefont {Novoselov},
  \citenamefont {Gorbachev}, \citenamefont {Shin}, \citenamefont {Fal'ko},\
  and\ \citenamefont {Tartakovskii}}]{Alexeev2019}%
  \BibitemOpen
  \bibfield  {author} {\bibinfo {author} {\bibfnamefont {E.~M.}\ \bibnamefont
  {Alexeev}}, \bibinfo {author} {\bibfnamefont {D.~A.}\ \bibnamefont
  {Ruiz-Tijerina}}, \bibinfo {author} {\bibfnamefont {M.}~\bibnamefont
  {Danovich}}, \bibinfo {author} {\bibfnamefont {M.~J.}\ \bibnamefont {Hamer}},
  \bibinfo {author} {\bibfnamefont {D.~J.}\ \bibnamefont {Terry}}, \bibinfo
  {author} {\bibfnamefont {P.~K.}\ \bibnamefont {Nayak}}, \bibinfo {author}
  {\bibfnamefont {S.}~\bibnamefont {Ahn}}, \bibinfo {author} {\bibfnamefont
  {S.}~\bibnamefont {Pak}}, \bibinfo {author} {\bibfnamefont {J.}~\bibnamefont
  {Lee}}, \bibinfo {author} {\bibfnamefont {J.~I.}\ \bibnamefont {Sohn}},
  \bibinfo {author} {\bibfnamefont {M.~R.}\ \bibnamefont {Molas}}, \bibinfo
  {author} {\bibfnamefont {M.}~\bibnamefont {Koperski}}, \bibinfo {author}
  {\bibfnamefont {K.}~\bibnamefont {Watanabe}}, \bibinfo {author}
  {\bibfnamefont {T.}~\bibnamefont {Taniguchi}}, \bibinfo {author}
  {\bibfnamefont {K.~S.}\ \bibnamefont {Novoselov}}, \bibinfo {author}
  {\bibfnamefont {R.~V.}\ \bibnamefont {Gorbachev}}, \bibinfo {author}
  {\bibfnamefont {H.~S.}\ \bibnamefont {Shin}}, \bibinfo {author}
  {\bibfnamefont {V.~I.}\ \bibnamefont {Fal'ko}}, \ and\ \bibinfo {author}
  {\bibfnamefont {A.~I.}\ \bibnamefont {Tartakovskii}},\ }\bibfield  {title}
  {\enquote {\bibinfo {title} {Resonantly hybridized excitons in moir{\'e}
  superlattices in van der waals heterostructures},}\ }\href {\doibase
  10.1038/s41586-019-0986-9} {\bibfield  {journal} {\bibinfo  {journal}
  {Nature}\ }\textbf {\bibinfo {volume} {567}},\ \bibinfo {pages} {81--86}
  (\bibinfo {year} {2019})}\BibitemShut {NoStop}%
\bibitem [{\citenamefont {Geim}(2013)}]{Geim2013}%
  \BibitemOpen
  \bibfield  {author} {\bibinfo {author} {\bibfnamefont {I.~V.}\ \bibnamefont
  {Geim}, \bibfnamefont {A.~K.and~Grigorieva}},\ }\bibfield  {title} {\enquote
  {\bibinfo {title} {Van der waals heterostructures},}\ }\href {\doibase
  10.1038/nature12385} {\bibfield  {journal} {\bibinfo  {journal} {Nature}\
  }\textbf {\bibinfo {volume} {499}},\ \bibinfo {pages} {419--425} (\bibinfo
  {year} {2013})}\BibitemShut {NoStop}%
\bibitem [{\citenamefont {Kunstmann}\ \emph {et~al.}(2018)\citenamefont
  {Kunstmann}, \citenamefont {Mooshammer}, \citenamefont {Nagler},
  \citenamefont {Rodolfo}, \citenamefont {Chaves}, \citenamefont {Stein},
  \citenamefont {Paradiso}, \citenamefont {Plechinger}, \citenamefont {Strunk},
  \citenamefont {Schüller}, \citenamefont {Seifert}, \citenamefont
  {Reichman},\ and\ \citenamefont {Korn}}]{NatureJENS}%
  \BibitemOpen
  \bibfield  {author} {\bibinfo {author} {\bibfnamefont {J.}~\bibnamefont
  {Kunstmann}}, \bibinfo {author} {\bibfnamefont {F.}~\bibnamefont
  {Mooshammer}}, \bibinfo {author} {\bibfnamefont {P.}~\bibnamefont {Nagler}},
  \bibinfo {author} {\bibnamefont {Rodolfo}}, \bibinfo {author} {\bibfnamefont
  {A.}~\bibnamefont {Chaves}}, \bibinfo {author} {\bibfnamefont
  {F.}~\bibnamefont {Stein}}, \bibinfo {author} {\bibfnamefont
  {N.}~\bibnamefont {Paradiso}}, \bibinfo {author} {\bibfnamefont
  {G.}~\bibnamefont {Plechinger}}, \bibinfo {author} {\bibfnamefont
  {C.}~\bibnamefont {Strunk}}, \bibinfo {author} {\bibfnamefont
  {C.}~\bibnamefont {Schüller}}, \bibinfo {author} {\bibfnamefont
  {G.}~\bibnamefont {Seifert}}, \bibinfo {author} {\bibfnamefont {D.~R.}\
  \bibnamefont {Reichman}}, \ and\ \bibinfo {author} {\bibfnamefont
  {T.}~\bibnamefont {Korn}},\ }\bibfield  {title} {\enquote {\bibinfo {title}
  {Momentum-space indirect interlayer excitons in transition-metal
  dichalcogenide van der waals heterostructures},}\ }\href {\doibase
  10.1038/s41567-018-0123-y} {\bibfield  {journal} {\bibinfo  {journal} {Nature
  Physics}\ }\textbf {\bibinfo {volume} {14}} (\bibinfo {year} {2018}),\
  10.1038/s41567-018-0123-y}\BibitemShut {NoStop}%
\bibitem [{\citenamefont {Cao}\ \emph {et~al.}(2018{\natexlab{b}})\citenamefont
  {Cao}, \citenamefont {Fatemi}, \citenamefont {Fang}, \citenamefont
  {Watanabe}, \citenamefont {Taniguchi}, \citenamefont {Kaxiras},\ and\
  \citenamefont {Jarillo-Herrero}}]{Cao2018}%
  \BibitemOpen
  \bibfield  {author} {\bibinfo {author} {\bibfnamefont {Y.}~\bibnamefont
  {Cao}}, \bibinfo {author} {\bibfnamefont {V.}~\bibnamefont {Fatemi}},
  \bibinfo {author} {\bibfnamefont {S.}~\bibnamefont {Fang}}, \bibinfo {author}
  {\bibfnamefont {K.}~\bibnamefont {Watanabe}}, \bibinfo {author}
  {\bibfnamefont {T.}~\bibnamefont {Taniguchi}}, \bibinfo {author}
  {\bibfnamefont {E.}~\bibnamefont {Kaxiras}}, \ and\ \bibinfo {author}
  {\bibfnamefont {P.}~\bibnamefont {Jarillo-Herrero}},\ }\bibfield  {title}
  {\enquote {\bibinfo {title} {Unconventional superconductivity in magic-angle
  graphene superlattices},}\ }\href {\doibase 10.1038/nature26160} {\bibfield
  {journal} {\bibinfo  {journal} {Nature}\ }\textbf {\bibinfo {volume} {556}},\
  \bibinfo {pages} {43--50} (\bibinfo {year} {2018}{\natexlab{b}})}\BibitemShut
  {NoStop}%
\bibitem [{\citenamefont {Birowska}\ \emph {et~al.}(2019)\citenamefont
  {Birowska}, \citenamefont {Urban}, \citenamefont {Baranowski}, \citenamefont
  {Maude}, \citenamefont {Plochocka},\ and\ \citenamefont
  {Szwacki}}]{Birowska_2019}%
  \BibitemOpen
  \bibfield  {author} {\bibinfo {author} {\bibfnamefont {M.}~\bibnamefont
  {Birowska}}, \bibinfo {author} {\bibfnamefont {J.}~\bibnamefont {Urban}},
  \bibinfo {author} {\bibfnamefont {M.}~\bibnamefont {Baranowski}}, \bibinfo
  {author} {\bibfnamefont {D.~K.}\ \bibnamefont {Maude}}, \bibinfo {author}
  {\bibfnamefont {P.}~\bibnamefont {Plochocka}}, \ and\ \bibinfo {author}
  {\bibfnamefont {N.~G.}\ \bibnamefont {Szwacki}},\ }\bibfield  {title}
  {\enquote {\bibinfo {title} {The impact of hexagonal boron nitride
  encapsulation on the structural and vibrational properties of few layer black
  phosphorus},}\ }\href {\doibase 10.1088/1361-6528/ab0332} {\bibfield
  {journal} {\bibinfo  {journal} {Nanotechnology}\ }\textbf {\bibinfo {volume}
  {30}},\ \bibinfo {pages} {195201} (\bibinfo {year} {2019})}\BibitemShut
  {NoStop}%
\bibitem [{\citenamefont {Carr}\ \emph {et~al.}(2017)\citenamefont {Carr},
  \citenamefont {Massatt}, \citenamefont {Fang}, \citenamefont {Cazeaux},
  \citenamefont {Luskin},\ and\ \citenamefont {Kaxiras}}]{PhysRevB.95.075420}%
  \BibitemOpen
  \bibfield  {author} {\bibinfo {author} {\bibfnamefont {S.}~\bibnamefont
  {Carr}}, \bibinfo {author} {\bibfnamefont {D.}~\bibnamefont {Massatt}},
  \bibinfo {author} {\bibfnamefont {S.}~\bibnamefont {Fang}}, \bibinfo {author}
  {\bibfnamefont {P.}~\bibnamefont {Cazeaux}}, \bibinfo {author} {\bibfnamefont
  {M.}~\bibnamefont {Luskin}}, \ and\ \bibinfo {author} {\bibfnamefont
  {E.}~\bibnamefont {Kaxiras}},\ }\bibfield  {title} {\enquote {\bibinfo
  {title} {Twistronics: Manipulating the electronic properties of
  two-dimensional layered structures through their twist angle},}\ }\href
  {\doibase 10.1103/PhysRevB.95.075420} {\bibfield  {journal} {\bibinfo
  {journal} {Phys. Rev. B}\ }\textbf {\bibinfo {volume} {95}},\ \bibinfo
  {pages} {075420} (\bibinfo {year} {2017})}\BibitemShut {NoStop}%
\bibitem [{\citenamefont {Bukharaev}\ \emph {et~al.}(2018)\citenamefont
  {Bukharaev}, \citenamefont {Zvezdin}, \citenamefont {Pyatakov},\ and\
  \citenamefont {Fetisov}}]{Bukharaev2008}%
  \BibitemOpen
  \bibfield  {author} {\bibinfo {author} {\bibfnamefont {A.~A.}\ \bibnamefont
  {Bukharaev}}, \bibinfo {author} {\bibfnamefont {A.~K.}\ \bibnamefont
  {Zvezdin}}, \bibinfo {author} {\bibfnamefont {A.~P.}\ \bibnamefont
  {Pyatakov}}, \ and\ \bibinfo {author} {\bibfnamefont {Y.~K.}\ \bibnamefont
  {Fetisov}},\ }\bibfield  {title} {\enquote {\bibinfo {title} {Straintronics:
  a new trend in micro- and nanoelectronics and materials science},}\ }\href
  {\doibase 10.3367/UFNr.2018.01.038279} {\bibfield  {journal} {\bibinfo
  {journal} {Physics-Uspekhi}\ }\textbf {\bibinfo {volume} {61}},\ \bibinfo
  {pages} {1175--1212} (\bibinfo {year} {2018})}\BibitemShut {NoStop}%
\bibitem [{\citenamefont {Hu}, \citenamefont {Zhou},\ and\ \citenamefont
  {Dong}(2017)}]{C7CP03558F}%
  \BibitemOpen
  \bibfield  {author} {\bibinfo {author} {\bibfnamefont {T.}~\bibnamefont
  {Hu}}, \bibinfo {author} {\bibfnamefont {J.}~\bibnamefont {Zhou}}, \ and\
  \bibinfo {author} {\bibfnamefont {J.}~\bibnamefont {Dong}},\ }\bibfield
  {title} {\enquote {\bibinfo {title} {Strain induced new phase and
  indirect–direct band gap transition of monolayer inse},}\ }\href {\doibase
  10.1039/C7CP03558F} {\bibfield  {journal} {\bibinfo  {journal} {Phys. Chem.
  Chem. Phys.}\ }\textbf {\bibinfo {volume} {19}},\ \bibinfo {pages}
  {21722--21728} (\bibinfo {year} {2017})}\BibitemShut {NoStop}%
\bibitem [{\citenamefont {Tamleh}, \citenamefont {Rezaei},\ and\ \citenamefont
  {Jalilian}(2018)}]{TAMLEH2018339}%
  \BibitemOpen
  \bibfield  {author} {\bibinfo {author} {\bibfnamefont {S.}~\bibnamefont
  {Tamleh}}, \bibinfo {author} {\bibfnamefont {G.}~\bibnamefont {Rezaei}}, \
  and\ \bibinfo {author} {\bibfnamefont {J.}~\bibnamefont {Jalilian}},\
  }\bibfield  {title} {\enquote {\bibinfo {title} {Stress and strain effects on
  the electronic structure and optical properties of scn monolayer},}\ }\href
  {\doibase https://doi.org/10.1016/j.physleta.2017.11.025} {\bibfield
  {journal} {\bibinfo  {journal} {Physics Letters A}\ }\textbf {\bibinfo
  {volume} {382}},\ \bibinfo {pages} {339 -- 345} (\bibinfo {year}
  {2018})}\BibitemShut {NoStop}%
\bibitem [{\citenamefont {Milowska}, \citenamefont {Birowska},\ and\
  \citenamefont {Majewski}(2011)}]{AIPMilowska}%
  \BibitemOpen
  \bibfield  {author} {\bibinfo {author} {\bibfnamefont {K.}~\bibnamefont
  {Milowska}}, \bibinfo {author} {\bibfnamefont {M.}~\bibnamefont {Birowska}},
  \ and\ \bibinfo {author} {\bibfnamefont {J.~A.}\ \bibnamefont {Majewski}},\
  }\bibfield  {title} {\enquote {\bibinfo {title} {Mechanical, electrical, and
  magnetic properties of functionalized carbon nanotubes},}\ }\href {\doibase
  10.1063/1.3666632} {\bibfield  {journal} {\bibinfo  {journal} {AIP Conference
  Proceedings}\ }\textbf {\bibinfo {volume} {1399}},\ \bibinfo {pages}
  {827--828} (\bibinfo {year} {2011})}\BibitemShut {NoStop}%
\bibitem [{\citenamefont {Xue}\ \emph {et~al.}(2011)\citenamefont {Xue},
  \citenamefont {Sanchez-Yamagishi}, \citenamefont {Bulmash}, \citenamefont
  {Jacquod}, \citenamefont {Deshpande}, \citenamefont {Watanabe}, \citenamefont
  {Taniguchi}, \citenamefont {Jarillo-Herrero},\ and\ \citenamefont
  {LeRoy}}]{Xue2011}%
  \BibitemOpen
  \bibfield  {author} {\bibinfo {author} {\bibfnamefont {J.}~\bibnamefont
  {Xue}}, \bibinfo {author} {\bibfnamefont {J.}~\bibnamefont
  {Sanchez-Yamagishi}}, \bibinfo {author} {\bibfnamefont {D.}~\bibnamefont
  {Bulmash}}, \bibinfo {author} {\bibfnamefont {P.}~\bibnamefont {Jacquod}},
  \bibinfo {author} {\bibfnamefont {A.}~\bibnamefont {Deshpande}}, \bibinfo
  {author} {\bibfnamefont {K.}~\bibnamefont {Watanabe}}, \bibinfo {author}
  {\bibfnamefont {T.}~\bibnamefont {Taniguchi}}, \bibinfo {author}
  {\bibfnamefont {P.}~\bibnamefont {Jarillo-Herrero}}, \ and\ \bibinfo {author}
  {\bibfnamefont {B.~J.}\ \bibnamefont {LeRoy}},\ }\bibfield  {title} {\enquote
  {\bibinfo {title} {Scanning tunnelling microscopy and spectroscopy of
  ultra-flat graphene on hexagonal boron nitride},}\ }\href {\doibase
  10.1038/nmat2968} {\bibfield  {journal} {\bibinfo  {journal} {Nature
  Materials}\ }\textbf {\bibinfo {volume} {10}},\ \bibinfo {pages} {282--285}
  (\bibinfo {year} {2011})}\BibitemShut {NoStop}%
\bibitem [{\citenamefont {Kresse}\ and\ \citenamefont
  {M{\"u}ller}(1996)}]{KRESSE199615}%
  \BibitemOpen
  \bibfield  {author} {\bibinfo {author} {\bibfnamefont {G.}~\bibnamefont
  {Kresse}}\ and\ \bibinfo {author} {\bibfnamefont {J.~F.}\ \bibnamefont
  {M{\"u}ller}},\ }\bibfield  {title} {\enquote {\bibinfo {title} {Efficiency
  of ab-initio total energy calculations for metals and semiconductors using a
  plane-wave basis set},}\ }\href {\doibase
  http://dx.doi.org/10.1016/0927-0256(96)00008-0} {\bibfield  {journal}
  {\bibinfo  {journal} {Computational Materials Science}\ }\textbf {\bibinfo
  {volume} {6}},\ \bibinfo {pages} {15 -- 50} (\bibinfo {year}
  {1996})}\BibitemShut {NoStop}%
\bibitem [{\citenamefont {Kresse}\ and\ \citenamefont
  {Hafner}(1993)}]{PhysRevB.47.558}%
  \BibitemOpen
  \bibfield  {author} {\bibinfo {author} {\bibfnamefont {G.}~\bibnamefont
  {Kresse}}\ and\ \bibinfo {author} {\bibfnamefont {J.}~\bibnamefont
  {Hafner}},\ }\bibfield  {title} {\enquote {\bibinfo {title} {Ab initio},}\
  }\href {\doibase 10.1103/PhysRevB.47.558} {\bibfield  {journal} {\bibinfo
  {journal} {Phys. Rev. B}\ }\textbf {\bibinfo {volume} {47}},\ \bibinfo
  {pages} {558--561} (\bibinfo {year} {1993})}\BibitemShut {NoStop}%
\bibitem [{\citenamefont {Giannozzi}\ and\ \citenamefont {Stefano}(2009)}]{QE}%
  \BibitemOpen
  \bibfield  {author} {\bibinfo {author} {\bibfnamefont {P.}~\bibnamefont
  {Giannozzi}}\ and\ \bibinfo {author} {\bibfnamefont {B.}~\bibnamefont
  {Stefano}},\ }\bibfield  {title} {\enquote {\bibinfo {title} {Quantum
  espresso: a modular and open-source software project for quantum simulations
  of materials},}\ }\href@noop {} {\bibfield  {journal} {\bibinfo  {journal}
  {Journal of Physics: Condensed Matter}\ }\textbf {\bibinfo {volume} {21}},\
  \bibinfo {pages} {395502} (\bibinfo {year} {2009})}\BibitemShut {NoStop}%
\bibitem [{\citenamefont {Soler}\ \emph {et~al.}(2002)\citenamefont {Soler},
  \citenamefont {Artacho}, \citenamefont {Gale}, \citenamefont {Garc{\'{\i}}a},
  \citenamefont {Junquera}, \citenamefont {Ordej{\'{o}}n},\ and\ \citenamefont
  {S{\'{a}}nchez-Portal}}]{Soler_2002}%
  \BibitemOpen
  \bibfield  {author} {\bibinfo {author} {\bibfnamefont {J.~M.}\ \bibnamefont
  {Soler}}, \bibinfo {author} {\bibfnamefont {E.}~\bibnamefont {Artacho}},
  \bibinfo {author} {\bibfnamefont {J.~D.}\ \bibnamefont {Gale}}, \bibinfo
  {author} {\bibfnamefont {A.}~\bibnamefont {Garc{\'{\i}}a}}, \bibinfo {author}
  {\bibfnamefont {J.}~\bibnamefont {Junquera}}, \bibinfo {author}
  {\bibfnamefont {P.}~\bibnamefont {Ordej{\'{o}}n}}, \ and\ \bibinfo {author}
  {\bibfnamefont {D.}~\bibnamefont {S{\'{a}}nchez-Portal}},\ }\bibfield
  {title} {\enquote {\bibinfo {title} {The {SIESTA} method forab
  initioorder-nmaterials simulation},}\ }\href {\doibase
  10.1088/0953-8984/14/11/302} {\bibfield  {journal} {\bibinfo  {journal}
  {Journal of Physics: Condensed Matter}\ }\textbf {\bibinfo {volume} {14}},\
  \bibinfo {pages} {2745--2779} (\bibinfo {year} {2002})}\BibitemShut {NoStop}%
\bibitem [{\citenamefont {Momma}\ and\ \citenamefont {Izumi}(2008)}]{VESTA}%
  \BibitemOpen
  \bibfield  {author} {\bibinfo {author} {\bibfnamefont {K.}~\bibnamefont
  {Momma}}\ and\ \bibinfo {author} {\bibfnamefont {F.}~\bibnamefont {Izumi}},\
  }\bibfield  {title} {\enquote {\bibinfo {title} {{{\it VESTA}: a
  three-dimensional visualization system for electronic and structural
  analysis}},}\ }\href {\doibase 10.1107/S0021889808012016} {\bibfield
  {journal} {\bibinfo  {journal} {Journal of Applied Crystallography}\ }\textbf
  {\bibinfo {volume} {41}},\ \bibinfo {pages} {653--658} (\bibinfo {year}
  {2008})}\BibitemShut {NoStop}%
\bibitem [{\citenamefont {Smidstrup}\ \emph {et~al.}(2020)\citenamefont
  {Smidstrup}, \citenamefont {Markussen}, \citenamefont {Vancraeyveld},
  \citenamefont {Wellendorff}, \citenamefont {Schneider}, \citenamefont
  {Gunst}, \citenamefont {Verstichel}, \citenamefont {Stradi}, \citenamefont
  {Khomyakov}, \citenamefont {Vej-Hansen} \emph
  {et~al.}}]{smidstrup2020quantumatk}%
  \BibitemOpen
  \bibfield  {author} {\bibinfo {author} {\bibfnamefont {S.}~\bibnamefont
  {Smidstrup}}, \bibinfo {author} {\bibfnamefont {T.}~\bibnamefont
  {Markussen}}, \bibinfo {author} {\bibfnamefont {P.}~\bibnamefont
  {Vancraeyveld}}, \bibinfo {author} {\bibfnamefont {J.}~\bibnamefont
  {Wellendorff}}, \bibinfo {author} {\bibfnamefont {J.}~\bibnamefont
  {Schneider}}, \bibinfo {author} {\bibfnamefont {T.}~\bibnamefont {Gunst}},
  \bibinfo {author} {\bibfnamefont {B.}~\bibnamefont {Verstichel}}, \bibinfo
  {author} {\bibfnamefont {D.}~\bibnamefont {Stradi}}, \bibinfo {author}
  {\bibfnamefont {P.~A.}\ \bibnamefont {Khomyakov}}, \bibinfo {author}
  {\bibfnamefont {U.~G.}\ \bibnamefont {Vej-Hansen}},  \emph {et~al.},\
  }\bibfield  {title} {\enquote {\bibinfo {title} {Quantumatk: An integrated
  platform of electronic and atomic-scale modelling tools},}\ }\href@noop {}
  {\bibfield  {journal} {\bibinfo  {journal} {J. Phys: Condens. Matter}\
  }\textbf {\bibinfo {volume} {32}},\ \bibinfo {pages} {015901} (\bibinfo
  {year} {2020})}\BibitemShut {NoStop}%
\bibitem [{\citenamefont {Wolf}\ and\ \citenamefont {Yip}(1992)}]{taksiazka}%
  \BibitemOpen
  \bibfield  {author} {\bibinfo {author} {\bibfnamefont {D.}~\bibnamefont
  {Wolf}}\ and\ \bibinfo {author} {\bibfnamefont {S.}~\bibnamefont {Yip}},\
  }\href@noop {} {\emph {\bibinfo {title} {Materials Interfaces. Atomic-level
  structure and properties}}},\ \bibinfo {edition} {1st}\ ed.\ (\bibinfo
  {publisher} {Chapman \& Hall},\ \bibinfo {year} {1992})\BibitemShut {NoStop}%
\bibitem [{\citenamefont {{Virtanen}}\ \emph {et~al.}(2020)\citenamefont
  {{Virtanen}}, \citenamefont {{Gommers}}, \citenamefont {{Oliphant}},
  \citenamefont {{Haberland}}, \citenamefont {{Reddy}}, \citenamefont
  {{Cournapeau}}, \citenamefont {{Burovski}}, \citenamefont {{Peterson}},
  \citenamefont {{Weckesser}}, \citenamefont {{Bright}}, \citenamefont {{van
  der Walt}}, \citenamefont {{Brett}}, \citenamefont {{Wilson}}, \citenamefont
  {{Jarrod Millman}}, \citenamefont {{Mayorov}}, \citenamefont {{Nelson}},
  \citenamefont {{Jones}}, \citenamefont {{Kern}}, \citenamefont {{Larson}},
  \citenamefont {{Carey}}, \citenamefont {{Polat}}, \citenamefont {{Feng}},
  \citenamefont {{Moore}}, \citenamefont {{Vand erPlas}}, \citenamefont
  {{Laxalde}}, \citenamefont {{Perktold}}, \citenamefont {{Cimrman}},
  \citenamefont {{Henriksen}}, \citenamefont {{Quintero}}, \citenamefont
  {{Harris}}, \citenamefont {{Archibald}}, \citenamefont {{Ribeiro}},
  \citenamefont {{Pedregosa}}, \citenamefont {{van Mulbregt}},\ and\
  \citenamefont {{Contributors}}}]{SciPy}%
  \BibitemOpen
  \bibfield  {author} {\bibinfo {author} {\bibfnamefont {P.}~\bibnamefont
  {{Virtanen}}}, \bibinfo {author} {\bibfnamefont {R.}~\bibnamefont
  {{Gommers}}}, \bibinfo {author} {\bibfnamefont {T.~E.}\ \bibnamefont
  {{Oliphant}}}, \bibinfo {author} {\bibfnamefont {M.}~\bibnamefont
  {{Haberland}}}, \bibinfo {author} {\bibfnamefont {T.}~\bibnamefont
  {{Reddy}}}, \bibinfo {author} {\bibfnamefont {D.}~\bibnamefont
  {{Cournapeau}}}, \bibinfo {author} {\bibfnamefont {E.}~\bibnamefont
  {{Burovski}}}, \bibinfo {author} {\bibfnamefont {P.}~\bibnamefont
  {{Peterson}}}, \bibinfo {author} {\bibfnamefont {W.}~\bibnamefont
  {{Weckesser}}}, \bibinfo {author} {\bibfnamefont {J.}~\bibnamefont
  {{Bright}}}, \bibinfo {author} {\bibfnamefont {S.~J.}\ \bibnamefont {{van der
  Walt}}}, \bibinfo {author} {\bibfnamefont {M.}~\bibnamefont {{Brett}}},
  \bibinfo {author} {\bibfnamefont {J.}~\bibnamefont {{Wilson}}}, \bibinfo
  {author} {\bibfnamefont {K.}~\bibnamefont {{Jarrod Millman}}}, \bibinfo
  {author} {\bibfnamefont {N.}~\bibnamefont {{Mayorov}}}, \bibinfo {author}
  {\bibfnamefont {A.~R.~J.}\ \bibnamefont {{Nelson}}}, \bibinfo {author}
  {\bibfnamefont {E.}~\bibnamefont {{Jones}}}, \bibinfo {author} {\bibfnamefont
  {R.}~\bibnamefont {{Kern}}}, \bibinfo {author} {\bibfnamefont
  {E.}~\bibnamefont {{Larson}}}, \bibinfo {author} {\bibfnamefont
  {C.}~\bibnamefont {{Carey}}}, \bibinfo {author} {\bibfnamefont
  {{\.I}.}~\bibnamefont {{Polat}}}, \bibinfo {author} {\bibfnamefont
  {Y.}~\bibnamefont {{Feng}}}, \bibinfo {author} {\bibfnamefont {E.~W.}\
  \bibnamefont {{Moore}}}, \bibinfo {author} {\bibfnamefont {J.}~\bibnamefont
  {{Vand erPlas}}}, \bibinfo {author} {\bibfnamefont {D.}~\bibnamefont
  {{Laxalde}}}, \bibinfo {author} {\bibfnamefont {J.}~\bibnamefont
  {{Perktold}}}, \bibinfo {author} {\bibfnamefont {R.}~\bibnamefont
  {{Cimrman}}}, \bibinfo {author} {\bibfnamefont {I.}~\bibnamefont
  {{Henriksen}}}, \bibinfo {author} {\bibfnamefont {E.~A.}\ \bibnamefont
  {{Quintero}}}, \bibinfo {author} {\bibfnamefont {C.~R.}\ \bibnamefont
  {{Harris}}}, \bibinfo {author} {\bibfnamefont {A.~M.}\ \bibnamefont
  {{Archibald}}}, \bibinfo {author} {\bibfnamefont {A.~H.}\ \bibnamefont
  {{Ribeiro}}}, \bibinfo {author} {\bibfnamefont {F.}~\bibnamefont
  {{Pedregosa}}}, \bibinfo {author} {\bibfnamefont {P.}~\bibnamefont {{van
  Mulbregt}}}, \ and\ \bibinfo {author} {\bibfnamefont {S.~.~.}\ \bibnamefont
  {{Contributors}}},\ }\bibfield  {title} {\enquote {\bibinfo {title} {{SciPy
  1.0: Fundamental Algorithms for Scientific Computing in Python}},}\ }\href
  {\doibase https://doi.org/10.1038/s41592-019-0686-2} {\bibfield  {journal}
  {\bibinfo  {journal} {Nature Methods}\ }\textbf {\bibinfo {volume} {17}},\
  \bibinfo {pages} {261--272} (\bibinfo {year} {2020})}\BibitemShut {NoStop}%
\bibitem [{\citenamefont {pandas~development team}(2020)}]{pandas}%
  \BibitemOpen
  \bibfield  {author} {\bibinfo {author} {\bibfnamefont {T.}~\bibnamefont
  {pandas~development team}},\ }\href {\doibase 10.5281/zenodo.3509134}
  {\enquote {\bibinfo {title} {pandas-dev/pandas: Pandas},}\ } (\bibinfo {year}
  {2020})\BibitemShut {NoStop}%
\bibitem [{git(2020)}]{github}%
  \BibitemOpen
  \href@noop {} {}\bibinfo {howpublished} {\url{
  "https://github.com/tnecio/supercell-core"}} (\bibinfo {year} {2020}),\
  \bibinfo {note} {accessed 2020-03-11}\BibitemShut {NoStop}%
\bibitem [{\citenamefont {Trambly~de Laissardi{\`e}re}, \citenamefont {Mayou},\
  and\ \citenamefont {Magaud}(2010)}]{TramblydeLaissardire2010}%
  \BibitemOpen
  \bibfield  {author} {\bibinfo {author} {\bibfnamefont {G.}~\bibnamefont
  {Trambly~de Laissardi{\`e}re}}, \bibinfo {author} {\bibfnamefont
  {D.}~\bibnamefont {Mayou}}, \ and\ \bibinfo {author} {\bibfnamefont
  {L.}~\bibnamefont {Magaud}},\ }\bibfield  {title} {\enquote {\bibinfo {title}
  {Localization of dirac electrons in rotated graphene bilayers},}\ }\href
  {\doibase 10.1021/nl902948m} {\bibfield  {journal} {\bibinfo  {journal} {Nano
  Letters}\ }\textbf {\bibinfo {volume} {10}},\ \bibinfo {pages} {804--808}
  (\bibinfo {year} {2010})}\BibitemShut {NoStop}%
\bibitem [{Pyt(2020)}]{PythonTB}%
  \BibitemOpen
  \href@noop {} {}\bibinfo {howpublished} {\url{
  "https://www.physics.rutgers.edu/pythtb/index.html"}} (\bibinfo {year}
  {2020}),\ \bibinfo {note} {python Tight Binding (PythTB)}\BibitemShut
  {NoStop}%
\bibitem [{\citenamefont {Birowska}, \citenamefont {Milowska},\ and\
  \citenamefont {Majewski}(2011)}]{APP2011.Birowska}%
  \BibitemOpen
  \bibfield  {author} {\bibinfo {author} {\bibfnamefont {M.}~\bibnamefont
  {Birowska}}, \bibinfo {author} {\bibfnamefont {K.}~\bibnamefont {Milowska}},
  \ and\ \bibinfo {author} {\bibfnamefont {J.~A.}\ \bibnamefont {Majewski}},\
  }\bibfield  {title} {\enquote {\bibinfo {title} {Van {D}er {W}aals {D}ensity
  {F}unctionals for {G}raphene {L}ayers and {G}raphite},}\ }\href@noop {}
  {\bibfield  {journal} {\bibinfo  {journal} {Acta Physica Polonica A}\
  }\textbf {\bibinfo {volume} {120}},\ \bibinfo {pages} {845--848} (\bibinfo
  {year} {2011})}\BibitemShut {NoStop}%
\end{thebibliography}%


\providecommand{\noopsort}[1]{}\providecommand{\singleletter}[1]{#1}%
%

\end{document}